\documentclass[sigconf, screen]{acmart}
\renewcommand\footnotetextcopyrightpermission[1]{}
\usepackage{xcolor}
\usepackage{listings}
\lstdefinelanguage{blueprint}{
  keywords=[1]{EMPTY},
  keywordstyle=[1]\color{gray},
  keywords=[2]{BLUEPRINT, NAME, TYPE, VERSION, ENVIRONMENT, DEPEND, WORKDIR, CMD, FROM, RUN},
  keywordstyle=[2]\color{blue}\bfseries,
  keywords=[3]{DockerHub, Apt, PyPI, LOCAL, PYTHON},
  keywordstyle=[3]\color{red}\bfseries,
  sensitive=true,
  morecomment=[l]{\#},
  morestring=[b]",
  otherkeywords={--------------------------------------------------},
}
\usepackage{enumitem}
\usepackage{textcomp}
\usepackage{multirow}
\usepackage{stfloats}

\usepackage[ruled,vlined]{algorithm2e}
\SetArgSty{textnormal}

\setlist[description]{topsep=2pt, itemsep=4pt, parsep=0pt, leftmargin=0pt}
\setlist[enumerate]{topsep=0pt, itemsep=0pt, parsep=0pt}
\setlist[itemize]{topsep=0pt, itemsep=0pt, parsep=0pt}

\title[CIR: Lightweight Container Image for Cross-Platform Deployment]{CIR: Lightweight Container Image \\for Cross-Platform Deployment}
\author{Fengzhi Li, Xiaohui Peng, Qingru Xu, Qisong Shi, Tuo Zhou, Yongxuan Dai, Yifan Wang,\\
Ninghui Sun, Zhiwei Xu}
\affiliation{%
  \institution{Institute of Computing Technology, Chinese Academy of Sciences}
  \city{}
  \country{}
}
\affiliation{%
  \institution{University of Chinese Academy of Sciences}
  \city{}
  \country{}
}

\begin{document}
\begin{abstract}
In modern cloud and heterogeneous distributed infrastructures, container images are widely used as the deployment unit for machine learning applications. An image bundles the application with its entire platform-specific execution environment and can be directly launched into a container instance. However, this approach forces developers to build and maintain separate images for each target deployment platform. This limitation is particularly evident for widely used interpreted languages such as Python and R in data analytics and machine learning, where application code is inherently cross-platform, yet the runtime dependencies are highly platform-specific. With emerging computing paradigms such as sky computing and edge computing, which demand seamless workload migration and cross-platform deployment, traditional images not only introduce inefficiencies in storage and network usage, but also impose substantial burdens on developers, who must repeatedly craft and manage platform-specific builds.

To address these challenges, we propose a \emph{lazy-build} approach that defers platform-specific construction to the deployment stage, thus keeping the image itself cross-platform. To enable this, we introduce a new image format, \emph{CIR} (Container Intermediate Representation), together with its pre-builder and lazy-builder. CIR targets interpreted-language applications and only stores the \emph{identifiers} of the application's direct dependencies, leaving platform adaptation to the lazy-builder, which at deployment time assembles the actual dependencies into runnable containers. A single CIR can therefore be deployed across heterogeneous platforms while reducing image size by 95\% compared to conventional images that bundle all dependencies. In evaluation, CIR shortens build time by 76--86\%, reduces network usage by 44--50\%, and achieves deployment speeds 40--60\% faster than pulling pre-built images, outperforming state-of-the-art systems such as Docker, Buildah, and Apptainer.
\end{abstract}
\maketitle
\section{Introduction}

\begin{figure}[tb]
    \centering
    \includegraphics[width=\linewidth]{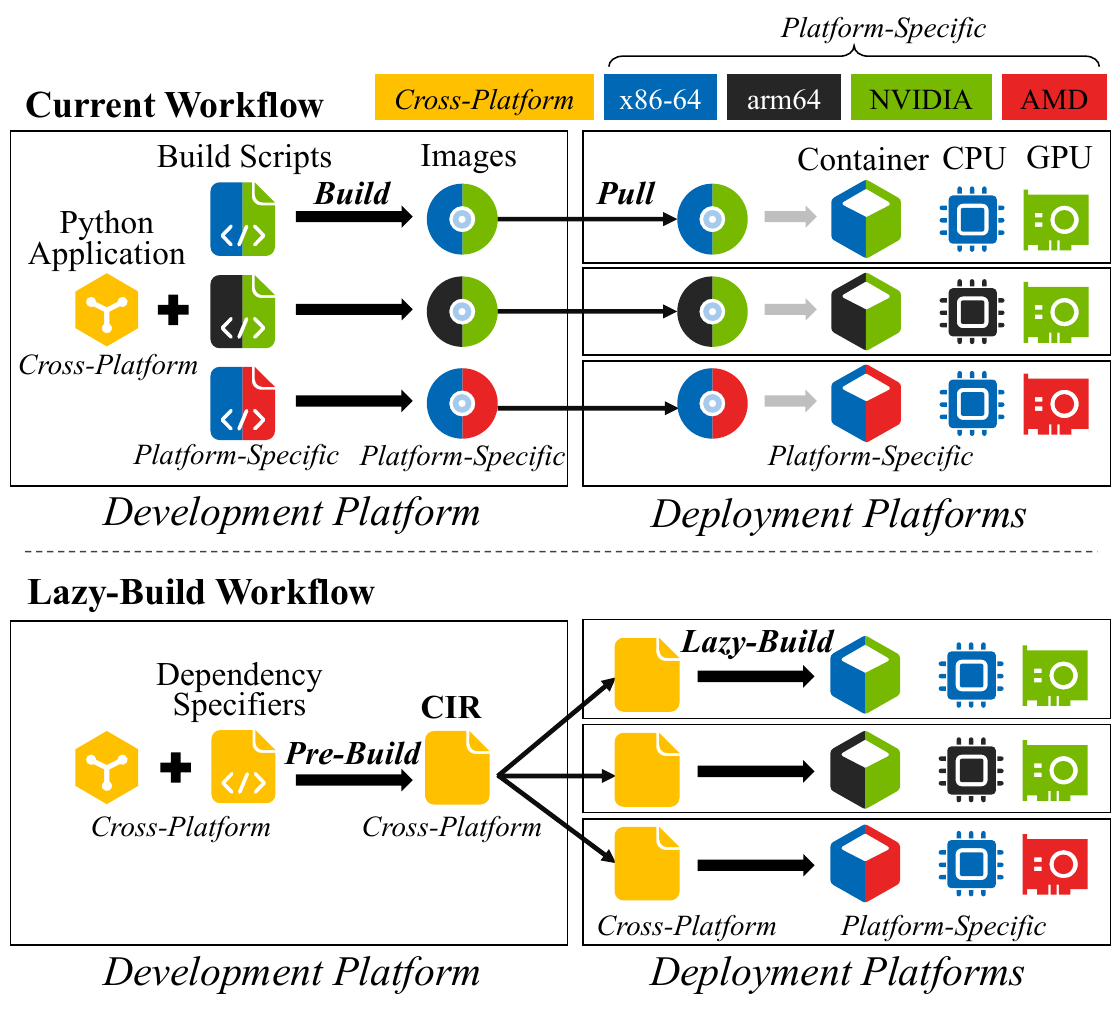}
    \caption{\textit{(Up)} Conventional container image is platform-specific, and multiple images have to be built for each CPU and GPU combination. \textit{(Down)} CIR is cross-platform, and a single CIR can be built into different container instances in different deployment platforms.}
    \label{fig:intro}
    \Description{}
    \vspace{-10pt}
\end{figure}

Over 90\% of cloud-native applications are deployed using containers~\cite{cncf2023survey}, an OS-level virtualization technology that provides isolated user-space environments. Applications are packaged together with their \emph{execution environments} into container images~\cite{dockerContainer}, which are stored in registries such as Docker Hub and retrieved by deployment platforms to instantiate containers.

Conventional container images are tightly coupled to specific deployment platforms, because the bundled execution environment contains libraries compiled for particular CPU or GPU architectures~\cite{HeterogeneousHardware}. The most commonly used image formats are Docker/OCI (produced by  \texttt{buildah} and \texttt{buildkit}) and SIF (produced by \texttt{Apptainer}).

In contrast, emerging distributed computing paradigms such as sky computing~\cite{skycomputing}, multi-cloud computing~\cite{multicloud}, and edge computing~\cite{edgecomputing} advocate \emph{cross-platform deployment} and \emph{workload migration} to enhance execution performance, data locality, and data security~\cite{edge1, edge2}. As shown in upper part of Figure~\ref{fig:intro}, the prevailing practice is to construct a separate image for each target deployment platform, which imposes substantial overhead on developers. For example, the YOLO11 object detection system~\cite{ultralytics} uses 10 distinct build scripts to generate images for different CPU-GPU combinations.

The root cause of this inefficiency lies in the separation of \emph{environment construction} and \emph{deployment platform information}. Environment construction is performed entirely on the developer side, whereas platform-specific details such as CPU architecture, GPU architecture, and driver versions are only visible to the container runtime at deployment time. This separation forces developers to anticipate the hardware characteristics of target deployment platforms and prepare dedicated build scripts. The image builder executes these scripts with environment managers such as \texttt{pip}, \texttt{apt}, \texttt{conda} and \texttt{nix} to create the execution environment and generate container images. At deployment, the container runtime detects hardware and driver information, selects the most compatible image, and launches containers. In short, developers build blind, while runtimes deploy blind, causing inefficiency and redundancy. This blind separation contradicts the design goals of modern parallel and distributed systems, where deployment decisions increasingly depend on runtime-visible hardware heterogeneity, accelerator availability, and data locality.

This mismatch is particularly pronounced for applications written in \emph{interpreted languages}, such as Python and R, which dominate data analytics and machine learning workloads. The application code of these languages is inherently cross-platform, while their execution environments heavily depend on platform-specific native libraries, GPU architectures, and driver versions. As a result, developers are forced to repeatedly construct and maintain platform-specific container images, even though the application logic itself remains unchanged.

\begin{figure}[tb]
    \centering
    \includegraphics[width=0.95\linewidth]{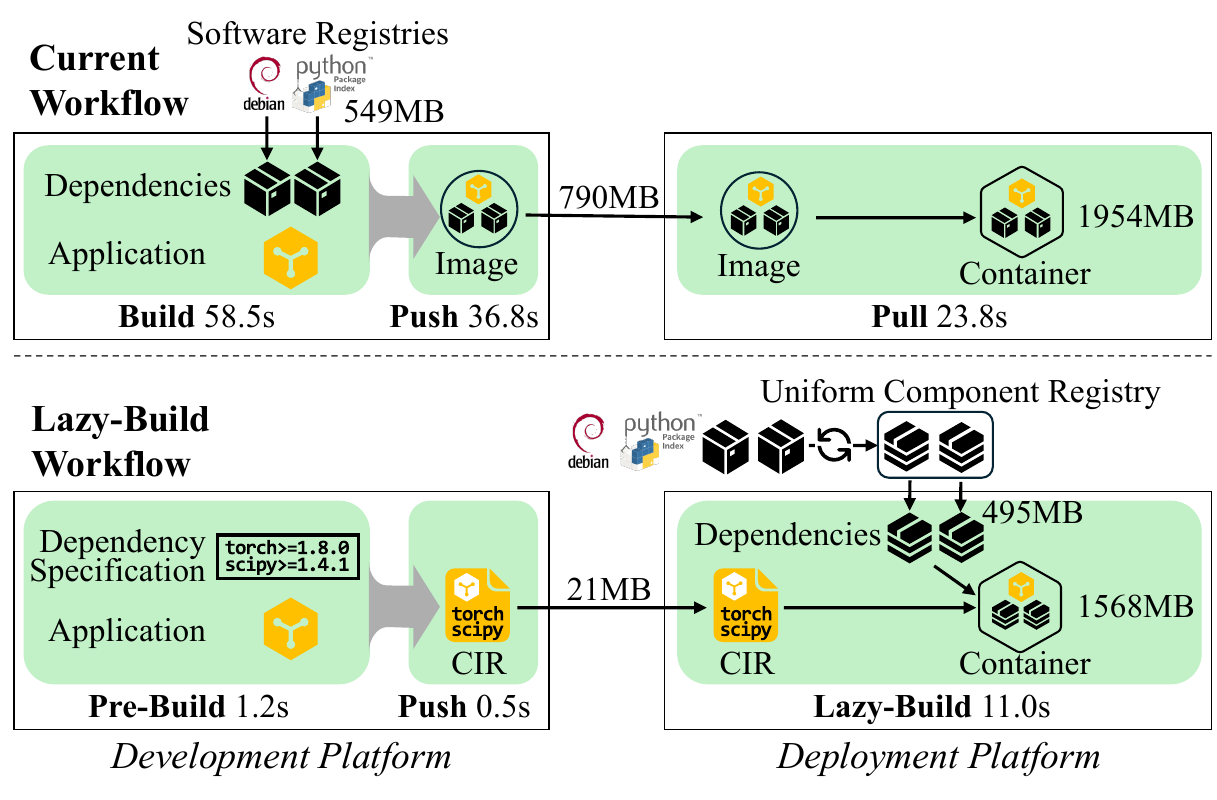}
    \caption{When deploying the YOLO11 project, the lazy-build approach shortens end-to-end deployment time from 199.1 s to 12.7 s and reduces network traffic from 1339 MB to 516 MB (measured on a 500 Mbps link).}
    \Description{}
    \label{fig:lazybuild}
\end{figure}

To bridge the gap between environment construction and deployment platform information, we propose a \emph{lazy-build method}, which defers the platform-specific portion of environment construction to the deployment platform. As shown in the lower part of Figure~\ref{fig:lazybuild}, the conventional \emph{build} and \emph{pull} phases are replaced with \emph{pre-build} and \emph{lazy-build} phases. The pre-builder on the development platform packages the cross-platform application code together with \emph{only the specifiers} of its direct dependencies into a new image format, the \emph{Container Intermediate Representation (CIR)}. At deployment time the lazy-builder performs three actions: (1) inspects the target hardware and driver configuration, (2) resolves the specifiers to concrete, platform-appropriate packages and binaries, and (3) assembles the selected packages with the application code into a runnable container instance.

The proposed CIR image format is inherently cross-platform, alleviating the burden on developers by abstracting away the complexity of heterogeneous hardware and software environments. Instead of manually crafting build scripts for each target platform, developers only need to declare the application’s direct dependencies. As shown in the lower part of Figure~\ref{fig:intro}, a single CIR can be lazily built into container instances across multiple platforms.

The CIR format is also lightweight, thus suitable for workload migration. Since current image bundles the entire execution environment, the average image size exceeds 1 GB on Docker Hub~\cite{imageSize}, incurring heavy overhead on network bandwidth~\cite{Starlight} and storage systems~\cite{Gear}. In contrast, CIR encodes only the names of application dependencies, reducing image size by up to 95\%.

The key challenges of this lazy-build method are ensuring \emph{correctness}, \emph{consistency}, and \emph{performance}. If the lazy-builder were to directly reuse existing \emph{environment managers} such as \texttt{pip} or \texttt{apt} on the deployment platform, the build process would be slower than pulling a pre-built image, and the results could be incorrect or inconsistent due to the instability of upstream software registries. To address these challenges, we introduce a new type of software package called the \emph{Uniform Component}, and co-design the container runtime, image builder, and environment manager into a unified lazy-builder.

The lazy-builder guarantees \emph{correctness} and \emph{consistency} by converting software packages into \emph{immutable} uniform components, and by applying a \emph{uniform dependency resolution algorithm} that deterministically selects the appropriate components and their dependencies. For the same deployment configuration, the lazy-builder always produces identical containers from a CIR, thereby ensuring stability and reproducibility. To support this process, we provide a web service that automatically converts Debian packages and Python packages into uniform components, and stores them in a \emph{Uniform Component Registry}, which currently contains more than 50,000 packages. 

In terms of \emph{performance}, lazy-builder not only outperforms state-of-the-art builders such as Docker, Buildah, and Apptainer by 77\%-87\%, but also achieves deployment speed faster than pulling pre-built images.

This lazy-build method is particularly well-suited to data science and machine learning workloads, which are predominantly developed by interpreted-languages. Our method enables a single image representation to be reused across heterogeneous CPU and GPU platforms, significantly reducing build overhead and accelerating deployment. For compiled languages, our system provides compatibility with existing container workflows; however, true cross-platform reuse is fundamentally constrained by architecture-specific binaries and is therefore outside the primary scope of this work.

This paper makes the following contributions:

\begin{itemize}
    \item We propose a lazy-build methodology for containers, that abstracts away heterogeneous platform specifics from developers, and enables architecture-aware optimizations during deployment-time.
    \item We design a lightweight cross-platform image format CIR, and develop a production-grade CIR build system that pre-build applications into CIRs and subsequently lazy-build CIRs into containers.
    \item We evaluate our proposed build system using nine real-world AI/ML applications across four different deployment platforms.
\end{itemize}

\section{Motivation}

Application migration across heterogeneous platforms is a key capability in emerging computing paradigms. In sky computing~\cite{skycomputing, skypilot, skyplane}, inter-cloud brokers migrate applications across heterogeneous platforms spanning multiple cloud providers, aiming to reduce costs, improve execution performance, and ensure regulatory compliance. Similarly, in edge computing~\cite{hcontainer, edgemigration}, applications are migrated from centralized cloud to distributed, heterogeneous nodes to minimize latency and alleviate data traffic. Even within a single cloud provider, platform diversity continues to grow~\cite{heterogeneouscloud}, driven by the adoption of novel computing architectures (e.g., TPU~\cite{tpu}, NPU~\cite{npu}), emerging GPU families (e.g., NVIDIA Jetson series~\cite{jetson}, AMD RDNA series~\cite{amd}), and alternative CPU architectures such as RISC-V. This growing platform diversity enables performance gains by migrating applications to platforms best suited to each workload.

In current practice, cross‑platform migrations rely on multiple-architecture image support: developers build separate container images for each target platform and annotate them with the \texttt{multi‑arch} label, allowing the container runtime to select and pull the appropriate image for the deployment platform. However, this approach introduces two key drawbacks:

\begin{description}
\item[Heavy developer burden] As shown in the upper part of Figure~\ref{fig:lazybuild}, the current workflow places the entire responsibility of image building on developers. To support diverse heterogeneous platforms, developers must gain extensive knowledge of platform-specific features and optimizations, create multiple image building scripts, build various container images, and thoroughly test each of them. Well-maintained projects highlight this overhead: YOLO11~\cite{ultralytics}, HuggingFace Transformers~\cite{transformers}, and TensorFlow~\cite{tensorflow} respectively maintain 10, 25, and 49 distinct Dockerfiles tailored for various CPU and GPU architectures. Similarly, projects such as PyTorch utilize Dockerfile templates with configurable parameters to generate different images suitable for multiple platforms. Both approaches significantly amplify the complexity and overall development effort, consequently imposing considerable additional maintenance burdens on developers.
\end{description}

Continuously rebuilding and updating container images is also essential for developers to address software bugs and vulnerabilities. Many applications employ CI/CD pipelines (such as Docker Build Cloud \footnote{\url{https://build.docker.com/}} and Google Cloud Build \footnote{\url{https://cloud.google.com/build/}}) to periodically regenerate and refresh container images. However, this practice necessitates significant extra efforts from developers, as well as incurs considerable computational resources, network bandwidth, and storage overhead. Consequently, a large portion of developers do not regularly update their container images, leaving Docker Hub community images with a median of 567 identified vulnerabilities~\cite{TechnicalLag}—over 82\% of which are classified as highly vulnerable due to outdated dependencies~\cite{Vulnerability}.

\begin{figure}[t]
    \centering
    \includegraphics[width=\linewidth]{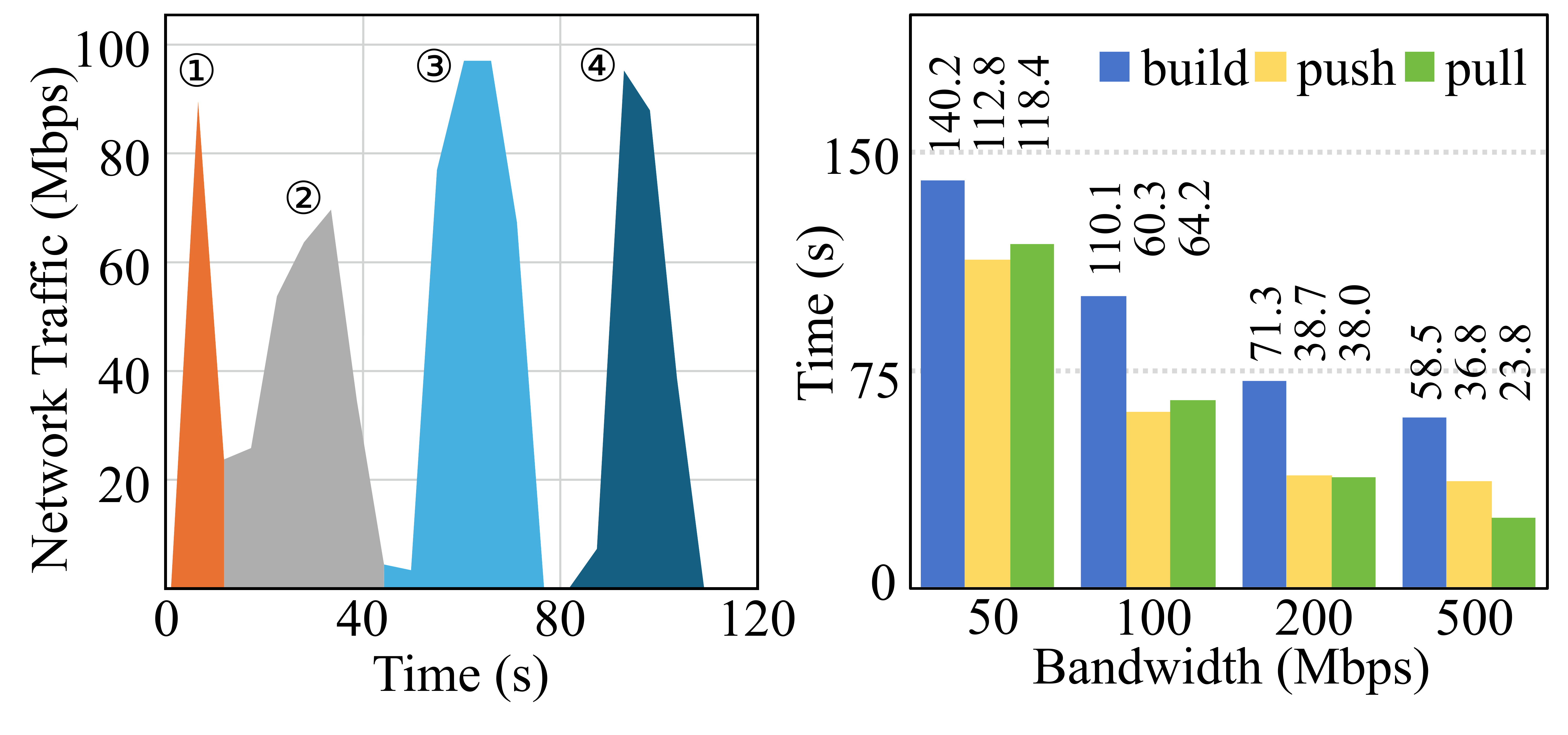}
    \caption{Building the image of YOLO11 project with Docker. \textit{(Left)} Shows the network traffic during the building process when the maximum bandwidth is set to 100 Mbps. \textit{(Right)} Shows the time usage of build, push and pull actions under different bandwidth limits.}
    \Description{}
    \vspace{-14pt}
    \label{fig:docker}
\end{figure}

\begin{description}
\item[Slow builds and deployments] To build a Docker image, developers first write a Dockerfile, which specifies a base image as the foundation of the building process and specifies the exact commands to use environment managers like \texttt{apt}, \texttt{conda} and \texttt{pip} for reproducing the execution environment. As shown in Figure~\ref{fig:uniform}, an image builder pulls the base image from a registry, launches a temporary container from that image, executes the Dockerfile’s instructions, invokes environment managers to install required software, and finally commits the resulting filesystem as a new image.
\end{description}

The performance of this image build process is throttled by the environment managers, which must run sequentially during each layer’s construction and cannot exploit cross‑manager parallelism. For example, the left side of Figure~\ref{fig:docker} plots network utilization while building the YOLO11 image over a 100 Mbps network. Regions 1-4 correspond to the four file layers: the base image, model weights, Debian packages, and Python libraries. In most cases, the network traffic achieves only partial utilization of the 100 Mbps capacity. The layers are built sequentially, with intervals between the different areas used for package installations.

Once the image is built, it can be pushed to the registry and later pulled on the deployment platform. Deployment latency is driven both by transfer volume and by the layer‑by‑layer unpacking process. As shown on the right side of Figure~\ref{fig:docker}, we measured pulling times for the 790 MB YOLO11 image under varying bandwidth caps. Given the image size, our calculations indicate that at low bandwidths, deployment is dominated by transfer time, whereas at high bandwidths, it is limited by the sequential unpacking of the file layers.

To address these limitations, we introduce CIR, a new image format that lets developers specify an application's dependencies once and reuse that single declaration across all target platforms, eliminating the need to create separate Dockerfiles manually. By encoding dependencies as abstract specifiers rather than embedding platform-specific binaries, CIR remains lightweight to distribute. To accelerate on-target assembly, we combine techniques such as container slimming, software component pre-processing, and parallelized dependency resolution and fetch. In our experiments this approach delivers up to 60\% faster deployments compared with conventional Docker-based workflows.

\section{Lazy-Build}

Lazy-build is the design principle of CIR's building system, separating the build process into two phases. On the development platform, the cross-platform application and its dependency specification are \emph{pre-built} into a CIR. Then on the deployment platform, the dependency specification guides the \emph{lazy-builder} to combine the actual required software components with the application into a container.

This section focuses on two research questions: (1) How should the CIR specify the dependencies in pre-build phase? (2) How to ensure the correctness and consistency of lazy-build? For the first question, we propose that CIR should specify \emph{declarative direct dependencies} (subsection~\ref{subsec:ddd}). For the second question, we propose to use immutable \emph{uniform components} as the container building blocks and design two algorithms to resolve the dependency (subsection~\ref{subsec:uniformcomponent}).

\subsection{Declarative Direct Dependencies}
\label{subsec:ddd}

\begin{figure}[tb]
    \centering
    \includegraphics[width=0.96\linewidth]{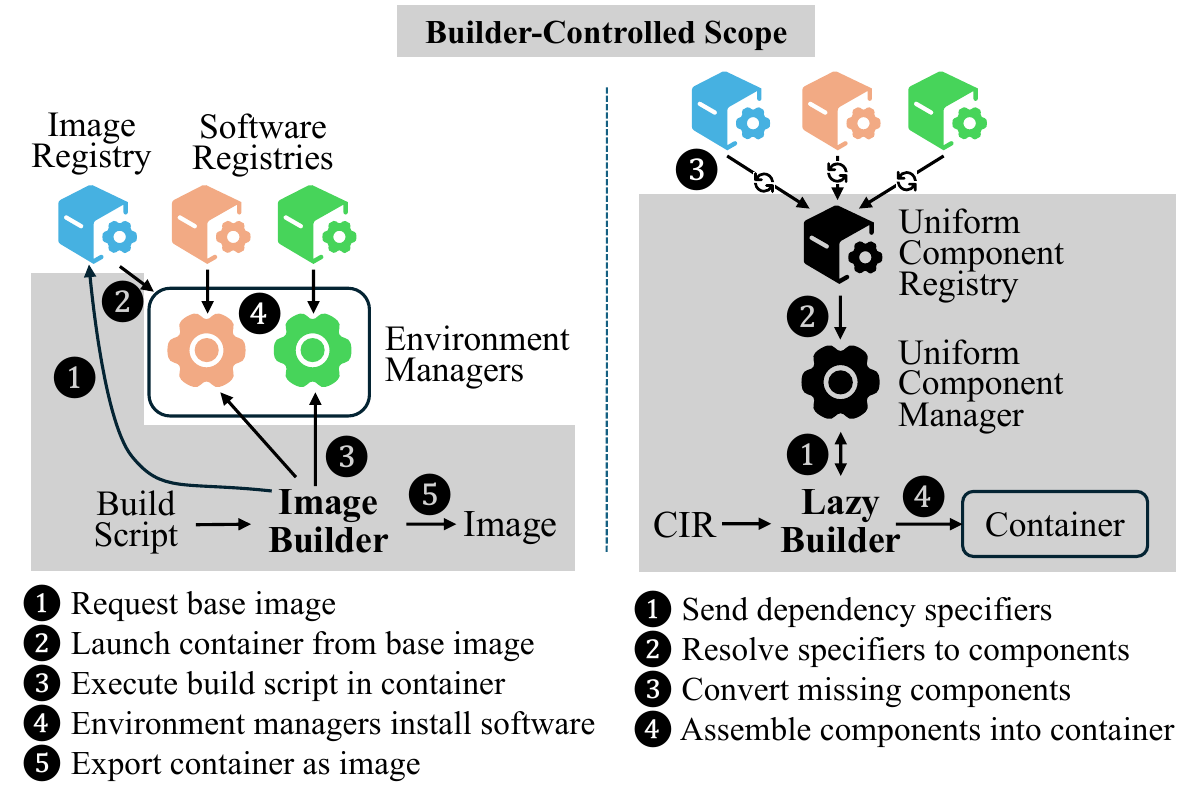}
    \caption{\textit{(Left)} Current image builder manages container content via environment managers. \textit{(Right)} Lazy-builder directly assembles containers from uniform components.}
    \Description{}
    \label{fig:uniform}
\end{figure}

Conventional image builders require developers to explicitly specify every build step. In contrast, CIR describes the execution environment solely through \emph{declarative direct dependencies}. This abstraction allows the lazy-builder on the deployment platform to interpret the declared dependencies and automatically select appropriate components, thereby constructing a platform-specific container instance on demand.

\begin{description}
\item[Declarative Dependency.] Current image builder is command-oriented rather than functionality-oriented. Consequently, environments with equivalent functionality may be created using different build scripts. For example, the following four Dockerfiles all build images that provide PyTorch.
\end{description}

\begin{lstlisting}[language=blueprint, numbers=none]
FROM debian:SOME_VERSION
RUN apt install -y python3-dev python3-pip
RUN pip install torch
--------------------------------------------------
FROM python:SOME_VERSION
RUN pip install torch
--------------------------------------------------
FROM nvidia/cuda:12.4.1-cudnn-runtime-ubuntu22.04
RUN apt install -y python3-dev python3-pip
RUN pip install torch
--------------------------------------------------
FROM pytorch/pytorch:2.5.1-cuda12.4-cudnn9-runtime
\end{lstlisting}

However, no single build method is universally applicable. Some methods produce slimmer images but place heavier demands on the host system, whereas others deliver better PyTorch performance at the cost of complicating the integration of additional functionalities. Ultimately, developers must rely on their own experience to select among these imperative methods for each platform configuration.

To solve this diversity problem, we propose to use a declarative method, declaring the needs of the application rather than specifying what to do during image build process. To include the PyTorch in the environment, the CIR should only declare
\begin{lstlisting}[language=blueprint, numbers=none]
    Python: torch == 2.5.1
\end{lstlisting}
and let the lazy-builder automatically select the most suitable build method on current platform. The detailed selection algorithm is described in section~\ref{subsec:uniformcomponent}.

\begin{description}
    \item[Direct Dependency.] Current image builders require developers to handle part of the indirect dependencies. For example, for OpenCV-Python functionality, the developers need to install system software dependencies including \texttt{libgl1-mesa-glx} and \texttt{libglib2.0-0}:
\end{description}
\begin{lstlisting}[language=blueprint, numbers=none]
FROM python:3.12-slim
RUN apt install -y libgl1-mesa-glx libglib2.0-0
RUN pip install opencv-python
\end{lstlisting}

For lazy-build method, the CIR only declares direct dependencies, delegating the resolution of all indirect dependencies to the lazy-builder. For instance, the developer only specifies \texttt{opencv-python == 4.9.0}, and the lazy-builder automatically resolves and retrieves the appropriate Python interpreter, \texttt{apt} packages, and \texttt{pip} packages. This approach not only improves deployment flexibility by allowing the lazy-builder to adapt dependencies to the deployment platform, but it also minimizes errors caused by developers specifying incorrect indirect dependencies.

To automatically resolve cross-manager dependency, it is critical that this approach centralizes the resolving process of indirect dependencies, managing them entirely through the builder. As shown in Figure~\ref{fig:uniform}, current image builders fetch base images from image registry and call multiple environment managers to install software and libraries. For lazy-build method, we propose to unify components from all registries, store them in one registry, and handle all their dependency relations in one component manager. Using this approach, we aim to render the deployment process more robust, efficient, and adaptive.

\subsection{Uniform Component}
\label{subsec:uniformcomponent}

To ensure correctness and consistency under lazy build, the lazy-builder maintains full control over the container execution environment. As shown in Figure~\ref{fig:uniform}, containers are constructed from uniform components, which serve as \emph{immutable} building blocks managed by a dedicated \emph{Uniform Component Manager}. These components are directly converted from existing software packages, enabling the lazy-builder to compose them through overlay without additional transformation, resulting in a functional container instance. 

Based on the declarative direct dependencies in CIR, the uniform component manager requires a selection algorithm to identify a suitable uniform component and a resolution algorithm to handle its indirect dependencies.

\begin{description}
    \item[Component Selection.] We identify a common process of component selection across environment managers for operating system software (\texttt{apt}, \texttt{nix}), programming language library (\texttt{pip}, \texttt{npm}, \texttt{conda}, \texttt{mamba}) and container image (\texttt{docker}). Based on this observation, we propose a uniform component selection algorithm (Algorithm~\ref{alg:selection}), which selects a specific uniform component according to a given dependency specification.
\end{description}

\begin{algorithm}[tb]
  \small
  \SetAlgoLined
  \KwIn{Dependency Item $d = (\mathcal{M}, n, \text{specifier})$}
  \KwOut{Uniform Component $c$}
  Initialize specSheet with host information\;
  $V \gets \text{VQ}(\mathcal{M}, n)$\;
  \Repeat{$e$ is not empty}{
    $v \gets \text{VS}_{\mathcal{M}}(V, \text{specifier})$\;
    \If{$v$ is empty} {
      \Return Error: no component satisfies \texttt{d}\;
    }
    $E \gets \text{EQ}(\mathcal{M}, n, v)$\;
    $e \gets \text{ES}_{\mathcal{M}}(E, \text{specSheet})$\;
    \If{$e$ is empty} {
      \tcc{current $v$ may not provide a suitable environment variant}
      $V \gets V \setminus v$\;
    }
  }
  $c \gets \text{CQ}(\mathcal{M}, n, v, e)$\;
  \caption{Uniform Component Selection}
  \label{alg:selection}
\end{algorithm}

In this algorithm, every component $c$ is uniquely identified by $(\mathcal{M}, n, v, e)$. For component manager $\mathcal{M}$, given a name $n$, there are many versions available. Then given the name $n$ and the version $v$, there are many variants built for different hardware and software environment $e$. Every \emph{Component Manager} $\mathcal{M}$ can be abstracted as two functions:
\begin{itemize}
    \item Version selection $\text{VS}_{\mathcal{M}}: (V, \text{specifier}) \rightarrow v_{\text{Best}}$
    \item Environment selection $\text{ES}_{\mathcal{M}}: (E, \text{specSheet}) \rightarrow e_{\text{Best}}$
\end{itemize}

$V$ is a set of versions, and VS selects the most suitable version $v_\text{Best}$ according to the specifier, which is a string like \texttt{>=3.0}, \texttt{\textasciitilde=2.0}, and \texttt{latest}.

$E$ represents a set of environment variants, and the ES function selects the most suitable variant, $e_\text{Best}$, based on the specSheet, which encapsulates the local hardware and software configurations. To rank the variants, we define a metric called \emph{deployability}, which measures the suitability of a component for deployment on the current platform. This metric considers factors such as local caching, component size, download time, and execution performance.


Each uniform component specifies its hardware and software requirements in its metadata. A \emph{deployability evaluator} reads the metadata of all candidates, compares their requirements against the specSheet, and calculates a deployability score. The ES function then selects the component with the highest deployability score.

A \emph{Uniform Component Registry} stores all the components for every manager, and provides three services:
\begin{itemize}
    \item Version query $\text{VQ}: (\mathcal{M}, n) \rightarrow V$
    \item Environment query $\text{EQ}: (\mathcal{M}, n, v) \rightarrow E$ 
    \item Component query $\text{CQ}: (\mathcal{M}, n, v, e) \rightarrow c$
\end{itemize}

A component selection request (for example the need of a Python library \texttt{torch} that satisfies \texttt{>=2.0}) corresponds to a dependency item $d = (\mathcal{M}, n, \text{specifier})$. It can be resolved to a specific component $c$ by the collaboration of component manager $\mathcal{M}$ and uniform component registry using Algorithm~\ref{alg:selection}.

\begin{description}
    \item[Dependency Resolution.] The uniform component selection algorithm leaves two issues unsolved: how is a specSheet generated, and how are indirect dependencies dealt with? We use a \emph{uniform dependency resolution algorithm} to address these issues, which is described in Algorithm~\ref{alg:resolution}.
\end{description}

\begin{algorithm}[t]
  \small
  \SetAlgoLined
  \KwIn{Application Dependencies $D$}
  \KwOut{Component List $L$}
  Initialize $\mathcal{C}$ with host information\;
  Initialize Dependency Tree \texttt{T}\;
  \tcc{Each node contains a dependency item $d$ and a resolution result component $c$}
  \texttt{T}.root $\gets$ (empty, $(D, \mathcal{C})$)\;
  \For{\texttt{dep} $\in D$} {
      \texttt{T}.root.AddChild((dep, empty))\;
  }
  \While{$T$ has a not resolved node}{
    \texttt{node} $\gets$ Breadth-First not resolved node of $T$\;
    \If{\texttt{node}.d.SatisfiedBy($L$)} {
      \textbf{continue}\;
    }
    \texttt{spec} $\gets $\texttt{node}.d.$\mathcal{M}$.getSpec($\mathcal{C}$)\;
    \Repeat{!\texttt{d}.hasConflict()}{
      \texttt{cs} $\gets$ \textbf{UniformComponentSelection}(\texttt{d}, \texttt{spec})\;
      \If{Selection fails}{
        \Return Error: no component satisfies \texttt{d}\;
      }
      \texttt{d} $\gets$ ConflictResolution(\texttt{T}, \texttt{cs})\;
    }
    \texttt{node}.c = \texttt{cs}\;
    \For{\texttt{dep} $\in$ \texttt{cs}.D} {
      \texttt{node}.AddChild((dep, empty))\;
    }
    $\mathcal{C} \gets$ CollectContext(\texttt{T})\;
    $L \gets$ CollectComponent(\texttt{T})\;
  }
  \caption{Uniform Dependency Resolution}
  \label{alg:resolution}
\end{algorithm}

This dependency resolution algorithm uses breadth-first order to generate the dependency tree, and use conflict-driven clause learning algorithm~\cite{Pubgrub} for conflict resolution. This algorithm is compatible with the dependency resolution algorithm in \texttt{apt} and \texttt{pip}, but also add a \emph{Building Context} $\mathcal{C}$ for every container building process for hardware-software compatibility and cross-manager compatibility. The building context is initialized by reading the hardware and software information of the host platform. For example,\\
\indent{\small
$\mathcal{C}_{\text{Init}} = \{\texttt{cpu}: \text{amd64},\ \texttt{gpu}:\text{nvidia},\ \texttt{gpu.driver}:535.183.01\}$
}

The metadata of uniform component $c = (D, \mathcal{C})$ includes dependency items $D$ and building context $C$. For example, OpenCV-Python component (\textbf{\texttt{pip}}, \texttt{opencv\_python}, \texttt{4.10.0.84}, \texttt{cp37-abi3- manylinux\_2\_17\_x86\_64}) is
{
\setlength{\jot}{1pt}
\begin{align*}
    c = ( D = \{ &(\textbf{\texttt{pip}}, \texttt{numpy}, \texttt{>=1.26.0}), \nonumber \\
    &(\textbf{\texttt{apt}}, \texttt{libgl1-mesa-glx}, \texttt{any}), \\
    &(\textbf{\texttt{apt}}, \texttt{libglib2.0-0}, \texttt{any})\}, \\
    \mathcal{C} = \{&\texttt{opencv.version}: 4.10.0.84\} )
\end{align*}
}

Each dependency item includes a component manager $\mathcal{M}$ item, which enables the expression of cross-manager dependencies.

\subsection{Correctness and Consistency}

The correctness of the build pipeline requires (1) accurate conversion of upstream software into uniform components, and (2) correct selection of those components according to the dependency specifiers. Using the image benchmark described in Section~\ref{subsec:benchmark}, we verified functional correctness by running the official functional tests for each application. To validate selection correctness, we compared the installed \texttt{apt} packages and \texttt{pip} packages in lazy-built containers against those produced by conventional image builders, confirming identical package versions and resolved artifacts.

We validate consistency by rebuilding each CIR across heterogeneous target platforms and comparing the results. After the first successful build, the lazy-builder is able to record the identifiers of all selected uniform components. Since uniform components are \emph{immutable}, an identical CIR deterministically produces bit-identical containers when rebuilt on the same platform type.
\section{CIR Build System}

\begin{figure*}[htb]
    \centering
    \includegraphics[width=0.95\linewidth]{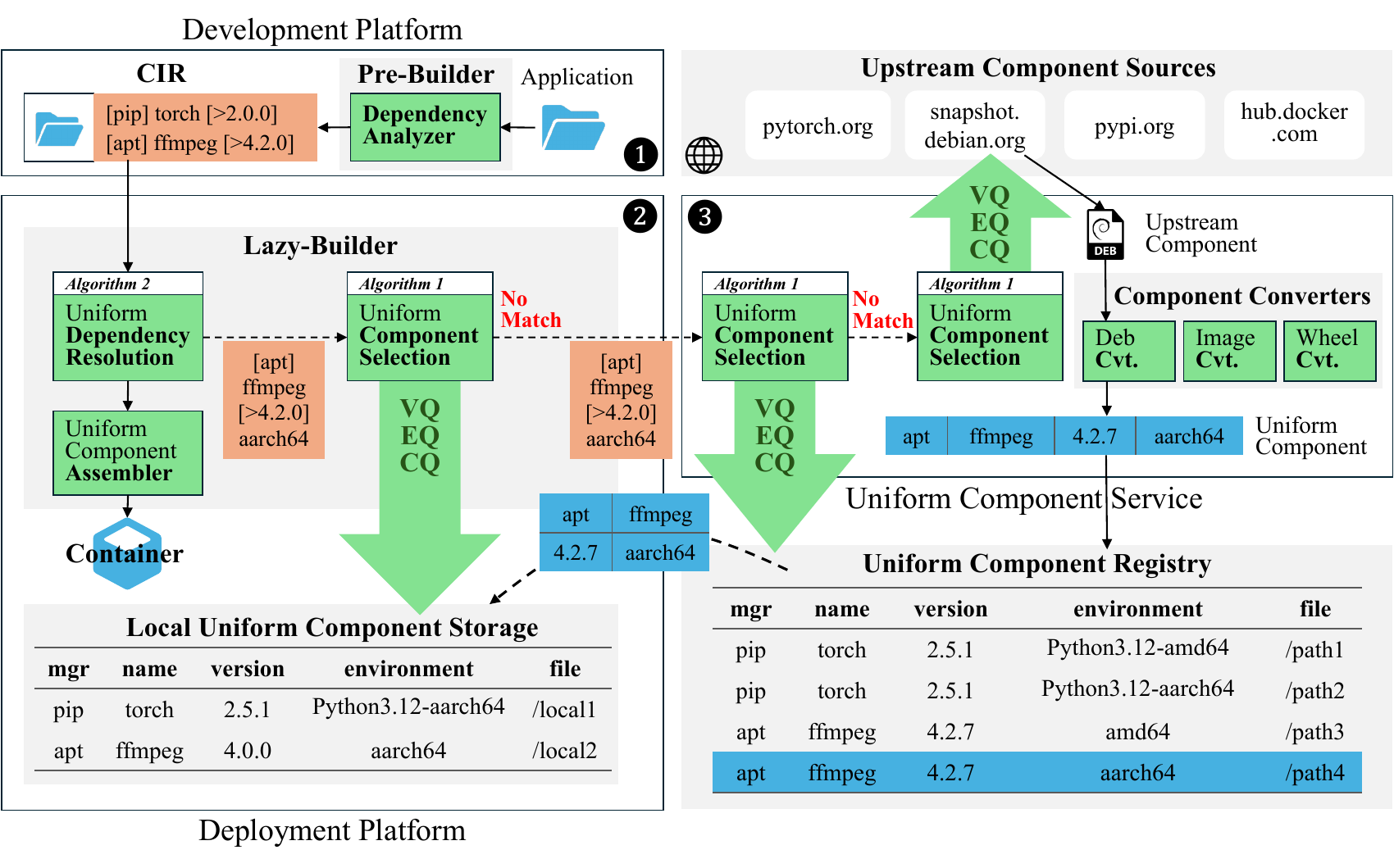}
    \vspace{-10pt}
    \caption{Overview of the CIR build system. \textit{(Upper Left)} On the development platform, the pre-builder packages the application into a CIR. \textit{(Lower Left)} On the deployment platform, the lazy-builder resolves dependencies described in the CIR and assembles uniform components into a container. If a required component is locally unavailable, the deployment platform sends the request to the uniform component service. \textit{(Right)} The service first attempts to resolve the request from the uniform component registry. If no suitable component exists, it fetches the component from upstream component sources and converts it into a uniform component.}
    \Description{}
    \label{fig:architecture}
\end{figure*}

Based on the proposed lazy-build method, we designed the CIR image format and implemented its build system with 13,000 LOC GoLang code, including a pre-builder (subsection~\ref{subsec:pre-builder}), a lazy-builder (subsection~\ref{subsec:lazy-builder}) and a uniform component service (subsection~\ref{subsec:service}).\footnote{\url{https://github.com/L-F-Z/TaskC}} Currently, this system supports Python projects and can automatically convert components from OCI images, Debian software packages, Python libraries, and HuggingFace models. Figure~\ref{fig:architecture} illustrates the architecture of our CIR build system.

\subsection{Pre-Builder}
\label{subsec:pre-builder}

The pre-builder on the development platform uses two methods for dependency analysis. The first method performs syntax analysis to identify dependencies across all code files in the Python application. The second method reads the analyzed results from other dependency analyzers, such as \texttt{requirements.txt} and \texttt{pyproject.toml}. Once the dependencies are collected, the pre-builder filters out the indirect dependencies and generates a metadata file, as shown below. The metadata, along with the application code, is then packed into a CIR.

\begin{lstlisting}[language=blueprint]
[NAME] YOLO11
[VERSION] 1.0
[DEPENDENCY]
- [Apt] libgl1 [any]
- [Apt] libglib2.0-0 [any]
- [Apt] libusb-1.0-0 [any]
- [PyPI] torch [>=1.8.0]
- [PyPI] torchvision [>=0.9.0]
...
- [LOCAL] /ultralytics [yolo11n.pt]
[WORKDIR] /ultralytics
\end{lstlisting}

\subsection{Lazy-Builder}
\label{subsec:lazy-builder}
The lazy-builder on the deployment platform reads a CIR, resolves all its dependencies, selects the suitable components, and assembles them into a container. The core of lazy-builder is the uniform component selection algorithm (Algorithm~\ref{alg:selection}) and the uniform dependency resolution algorithm (Algorithm~\ref{alg:resolution}). 

For Algorithm~\ref{alg:selection}, the VQ, EQ, and CQ functions are implemented by the \emph{Local Uniform Component Storage}, which caches components from uniform component service. The VS and ES rules are described in each component manager's specification document. For example, PyPA Specifications\footnote{\url{https://packaging.python.org/en/latest/specifications/}} describes the scheme of Python wheel version identifier, version specifier, platform tag, and platform compatibility. We followed these specifications to implement the VS and ES functions for selecting Python libraries and designed the specSheet for the Python component manager. The specSheet includes four parameters: CPU architecture, system type, Python interpreter version, and standard C library version. Similarly, we used Debian Policy Manual \footnote{\url{https://www.debian.org/doc/debian-policy/}} for Debian software packages and Docker Image Specification\footnote{\url{https://github.com/moby/docker-image-spec/}} for Docker images.

After all dependencies are resolved, and components are selected and cached, the \emph{Uniform Component Assembler} uses OverlayFS~\cite{overlayfs} to mount all the components into a root file system and generates an OCI runtime specification. containerd is used as the container runtime, with CNI plugins and NVIDIA Container Toolkit installed as network and GPU support. Since Docker also uses containerd as runtime, the CIR system achieves container startup latency and runtime performance identical to Docker.

Finally, upon the completion of deployment, the lazy-builder records the exact versions of all selected components and generates a dedicated version locking file for each platform. This file serves as a reproducibility manifest, ensuring consistent behavior across testing and production deployment platforms.

\subsection{Uniform Component Service}
\label{subsec:service}
Uniform component service provides uniform components for all platforms. Once it receives a request, the service first tries to select from the \emph{Uniform Component Registry} that stores all the previously converted uniform components. If none of them satisfies the request, the service then requests the \emph{Upstream Component Sources} for a component and uses \emph{Component Converters} to convert it into a uniform component. The core of this service comprises the uniform component selection algorithm and component converters.

The uniform component registry we implemented provides VQ, EQ, and PQ functions for Algorithm~\ref{alg:selection}'s queries. For upstream component sources, we created a portability layer to transform our functions to their API invocations (DockerHub API, Debian snapshot API and Python Package Index API.

Three component converters are used by the service to convert Docker images, Debian software packages, and Python libraries into uniform components, which are stored in a compressed \texttt{.tar.gz} archive format that supports streaming decompression. Each converter pre-compiles the component to simplify the installation process. Subsequently, the converter extracts the dependency information into a separate metadata file, allowing both dependency resolution and component downloading to be performed in parallel.

Docker image has a \texttt{manifest} file, which links to a \texttt{config} file and several \texttt{layer} files. Docker images normally do not exhibit dependency issues, so the converter only extracts metadata such as \texttt{User}, \texttt{WorkingDir}, \texttt{Env}, \texttt{Entrypoint} and \texttt{Cmd} from the \texttt{config} file. Multiple \texttt{layer} files are extracted to the same directory according to OCI image specification \footnote{\url{https://github.com/opencontainers/image-spec/}} to simulate the mounting result of OverlayFS~\cite{overlayfs}.

The Debian software package \texttt{.deb} file contains a \texttt{data.tar} file and a \texttt{control.tar} file. \texttt{data.tar} includes all the files that require unpacking. \texttt{control.tar} includes the dependency information, and scripts to be executed before and after the unpacking. The converter analyzes and simulates the execution of the scripts, including creating symbolic links, updating shared library caches, setting file permissions, and managing users. Dependency information is directly extracted to uniform component's metadata.

Python library binary distribution \texttt{.whl} contains a \texttt{.dist-info} directory and multiple files or directories that need to be placed in the Python \texttt{site-packages} directory. From the \texttt{METADATA} file in \texttt{.dist-info} directory, the converter extracts the dependency information and the Python version requirement to uniform component's metadata. The converter also creates executable scripts according to information in \texttt{entrypoint.txt}. For Python library source distribution, the converter uses Python 3.6 - 3.12 build tools to generate a \texttt{.whl} binary distribution for each version, and then converts the binary distributions to uniform components.

\section{Evaluation}
To assess the build performance and cross-platform capability, we compared the CIR build system with three state-of-the-art image builders, including \textbf{Docker} (version 27.3.1, using \textbf{buildx} as the frontend and \textbf{buildkit} as the backend build engine)~\cite{dockerContainer}, \textbf{Buildah} (version 1.28.2)~\cite{Buildah} and \textbf{Apptainer} (version 1.3.5)~\cite{apptainer}. We evaluated these builders on four different deployment platforms (subsection~\ref{subsec:setup}), using a benchmark suite (subsection~\ref{subsec:benchmark}) consisting of nine real-world representative machine learning applications. To demonstrate the cross-platform characteristics of CIR (subsection~\ref{subsec:charac}), we deployed the same CIR on the four deployment platforms.

The performance of build systems is commonly affected by three key factors: image size, computing resources, and network bandwidth. For image size, we compared the CIR with other image formats generated from baseline build systems (subsection~\ref{subsec:imagesize}). To emulate the network conditions (subsection~\ref{subsec:bandwidth}) of edge devices, personal computers, and data center servers, we adjusted the bandwidth between deployment platforms and the component server from 10 Mbps to 1000 Mbps. For computing resources (subsection~\ref{subsec:computing}), we tested from 1 CPU core with 2 GB memory to 16 CPU cores with 32 GB memory. We also evaluated the file-sharing performance (subsection~\ref{subsec:share}) of the CIR lazy-builder against state-of-the-art image de-duplication methods.

As shown in Figure~\ref{fig:lazybuild}, baseline systems perform build and push on the development platform and pull on the deployment platform. In contrast, CIR performs pre-build and push on the development platform and lazy-build on the deployment platform, which downloads the CIR image and assembles uniform components.

We evaluate both systems using three metrics: \textbf{Build Time} (build vs. pre-build + lazy-build), \textbf{Deployment Time} (pull vs. lazy-build), and \textbf{End-to-end Time} (build + push + pull vs. pre-build + push + lazy-build).

\subsection{Experimental Setup}
\label{subsec:setup}
We selected four representative platforms as deployment platforms, covering two CPU architectures (amd64 and aarch64) and two GPU architectures (NVIDIA Ampere and AMD RDNA 3). The hardware and software specifications are:
\begin{itemize}[leftmargin=10pt]
    \item \textbf{CPU Server}: Intel Xeon Platinum 8163, 64 GB Memory.
    \item \textbf{GPU Server}: Intel Xeon Gold 6442Y, 64 GB Memory. NVIDIA A100 80 GB PCIe. 535.183.01 Driver, CUDA 12.2.
    \item \textbf{Personal Computer}: Intel Core i5-10505, 16 GB Memory. AMD Radeon RX 7900 XTX 24 GB, 24.20.3 Driver.
    \item \textbf{Edge Device}: NVIDIA Jetson Orin NX 16 GB, Arm Cortex-A78AE CPU.
\end{itemize}

To eliminate the impact of external registry speeds, we used the same registry server to cache all external registries and host a uniform component registry. All deployment platforms are connected to this server via a network switch, which limits the maximum bandwidth to 1 Gbps.

\subsection{Benchmark Suite}
\label{subsec:benchmark}
Our benchmark\footnote{\url{https://github.com/L-F-Z/TaskC_Demos}} focuses on machine learning domain, which is extensively deployed across edge devices, personal computers, and data center servers. The benchmark suite encompasses nine real-world projects that cover key research areas in machine learning:
\begin{itemize}[leftmargin=10pt]
    \item \textbf{Computer Vision}: YOLO11~\cite{ultralytics}, SAM2~\cite{sam2}
    \item \textbf{Natural Language Processing}: Transformers~\cite{transformers}, LoRA~\cite{lora}
    \item \textbf{Multimodal}: CLIP~\cite{clip}, Stable Diffusion~\cite{stablediffusion}
    \item \textbf{Speech Processing}: TTS~\cite{tts}, Whisper~\cite{whisper}
    \item \textbf{Reinforcement Learning}: Stable Baselines 3~\cite{sb3}
\end{itemize}

For each project, we directly used the official Dockerfiles for Docker and Buildah, and authored Apptainer definition files that reproduce the same build steps for Apptainer.

\subsection{Cross-Platform CIR}
\label{subsec:charac}

We selected the YOLO11 project as our test case because the project’s repository includes multiple official Dockerfiles targeting different deployment platforms. We uses four distinct Dockerfiles to test on each platform, and the Docker builder used 348.9 s on personal computer, 416.1 s on edge device, 212.6 s on GPU server, and 111.2 s on CPU server.

In comparison, the CIR approach used a single CIR for all four deployment platforms. With this setup, the CIR build time was reduced by an average of \textbf{78.7\%}, which was only 118.4 s on a personal computer, 26.4 s on an edge device, 67.5 s on a GPU server, and 14.5 s on a CPU server.

\subsection{Image Size}
\label{subsec:imagesize}

\begin{figure}[htp]
    \centering
    \includegraphics[width=\linewidth]{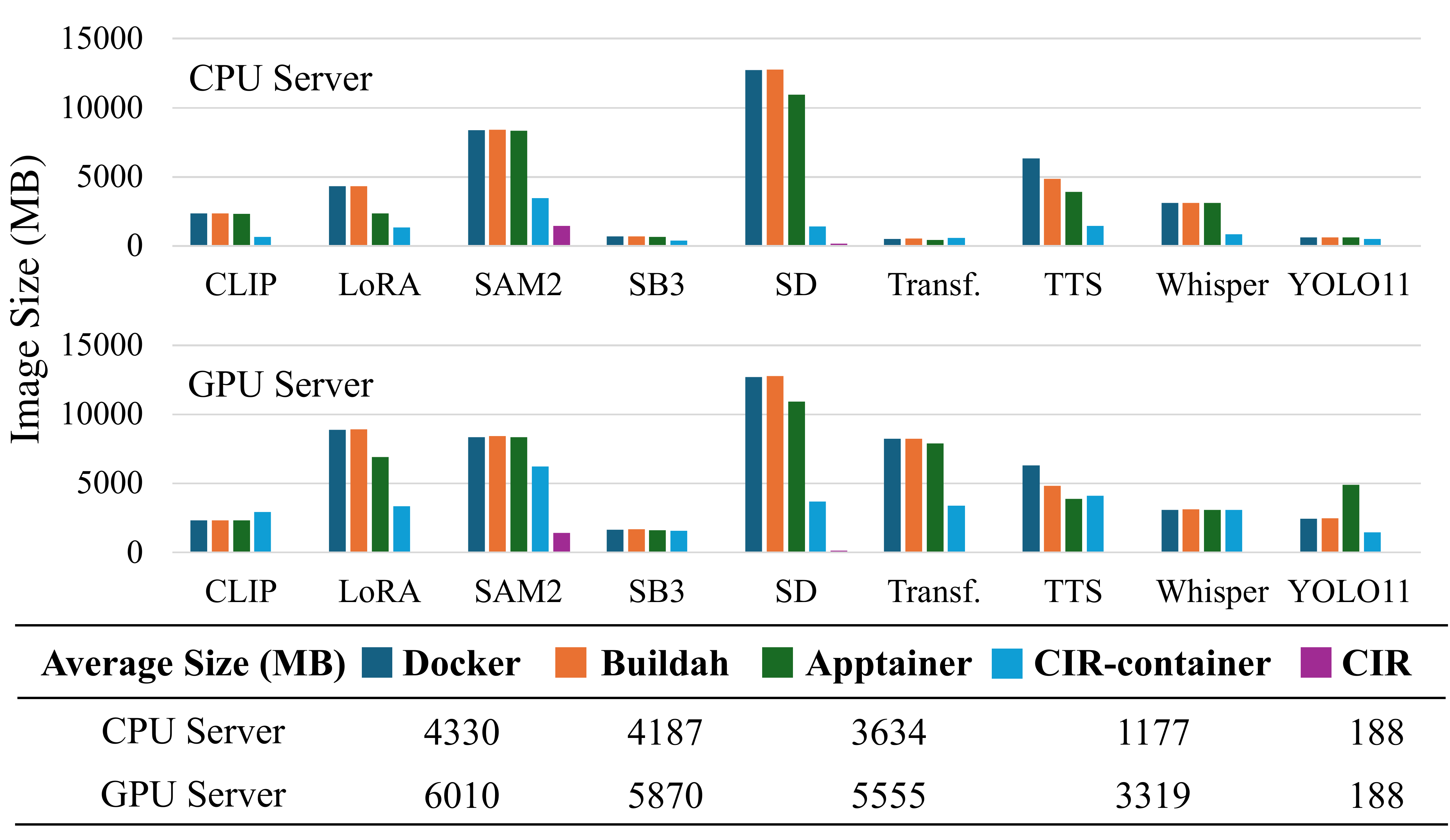}
    \captionsetup{type=figure}
    \caption{The image size for CPU and GPU platforms.}
    \label{fig:size}
\end{figure}

    


Next, we built the benchmark suite to assess image size generated by our CIR build system and baseline systems. The average image sizes generated on the GPU server by Docker, Buildah, and Apptainer were 6010 MB, 5870 MB, and 5555 MB, respectively, while the average CIR size was only 188 MB. Therefore, on the development platform, CIR reduces average data transfer by approximately 96\%. On the deployment platform, the lazy-builder averagely fetched 188 MB CIR and 1822 MB uniform component to assemble the full container, which was about 65\% less than Docker, Buildah, and Apptainer. This reduction in final container size is achieved by reusing GPU-related components on host platforms via \texttt{libnvidia-container}, omitting package managers such as \texttt{apt} and \texttt{pip} (since their tasks are performed by the lazy-builder), and removing package caches.

The CIR build system delivers better performance for two primary reasons: (1) CIR containers are smaller because they include only the necessary components, and (2) all components are pre-built and pre-processed, allowing for direct assembly. To demonstrate that the second factor also enhances performance, we used the CIR version locking mechanism to create nine specialized CIRs, referred to as CIR-locked, which specify exactly the same dependencies as the baseline images. The container sizes produced by CIR-locked are nearly identical to those of Buildah, deviating by no more than ±2\%. In the subsequent experiment, we used CIR-locked to confirm that even when container sizes are similar, the CIR build system still outperformed other baseline systems because of the pre-built and pre-processed components.

\subsection{Impact of Network Bandwidth}
\label{subsec:bandwidth}

\begin{figure}[htp]
    \centering
    \includegraphics[width=\linewidth]{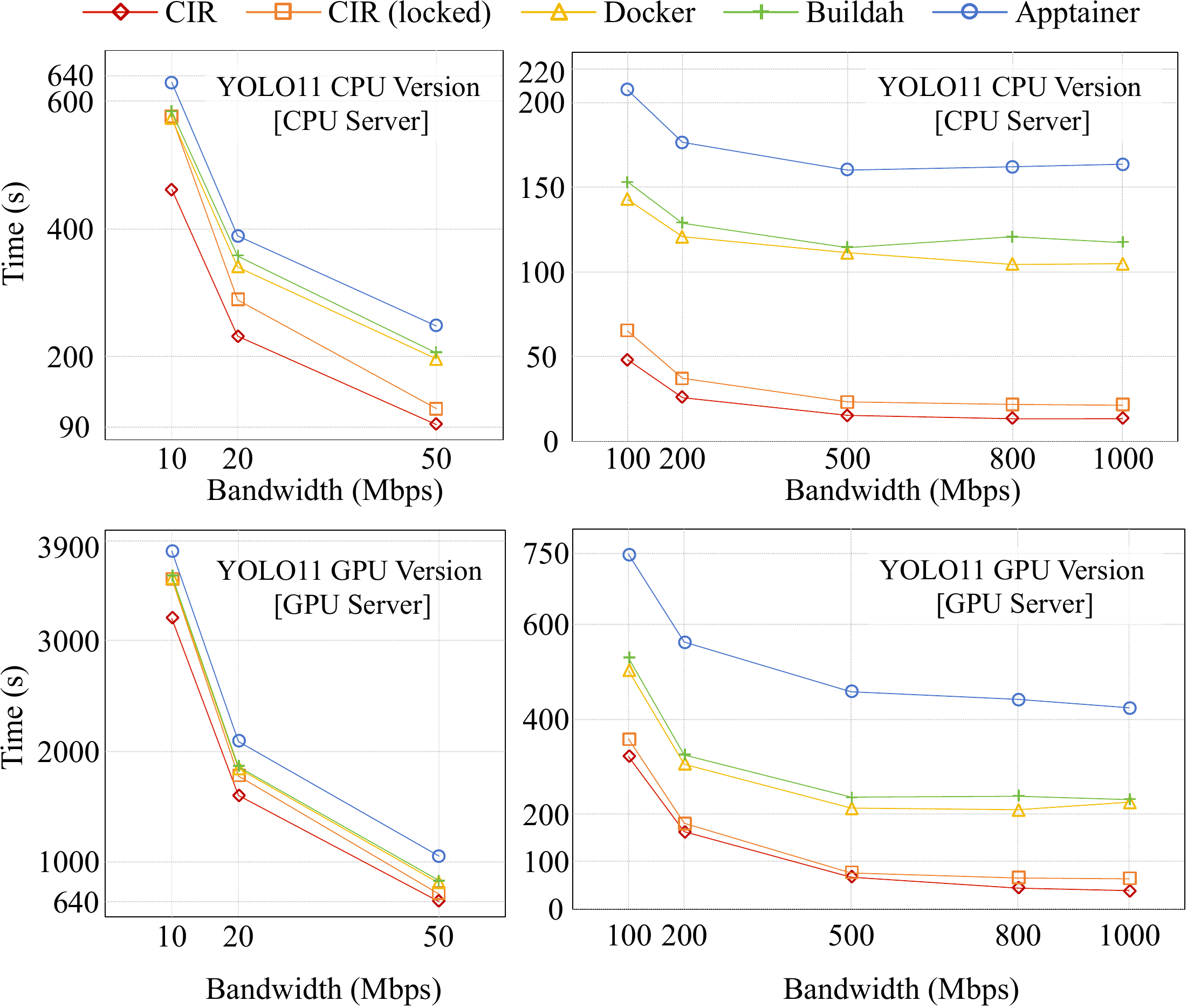}
    \captionsetup{type=figure}
    \caption{Build times of YOLO11 project under different network bandwidth.}
    \label{fig:bandwidth}
\end{figure}

To demonstrate the impact of network bandwidth, we built YOLO11 on both CPU and GPU servers while limiting the network bandwidth to 10, 20, 50, 100, 200, 500, 800, and 1000 Mbps. As shown in Figure~\ref{fig:bandwidth}, the CIR build system consistently outperforms all other systems at every tested bandwidth, achieving average build time reductions of 54.1\% compared to Docker, 56.0\% compared to Buildah, and 64.9\% compared to Apptainer.

Notably, the build time difference between Docker and the CIR system remained roughly 100 s across all bandwidth settings. This observation indicates that with increasing bandwidth, the relative advantage of CIR over Docker increases. When we used a 1000 Mbps network bandwidth, CIR system was 87.1\% faster than Docker. The persistent 100-second gap represents the time spent by the Docker on component compilation and processing, which cannot be reduced by simply increasing network bandwidth.

Because the CIR-locked container is similar in size to the Docker container but employs pre-compiled and pre-processed components, the consistent 100-second difference also explains why CIR-locked’s build time matches Docker’s under lower bandwidth conditions yet more closely approaches CIR’s performance as the available bandwidth increases.

\subsection{Impact of Computing Resources}
\label{subsec:computing}

\begin{figure}[ht]
    \centering
    \includegraphics[width=\linewidth]{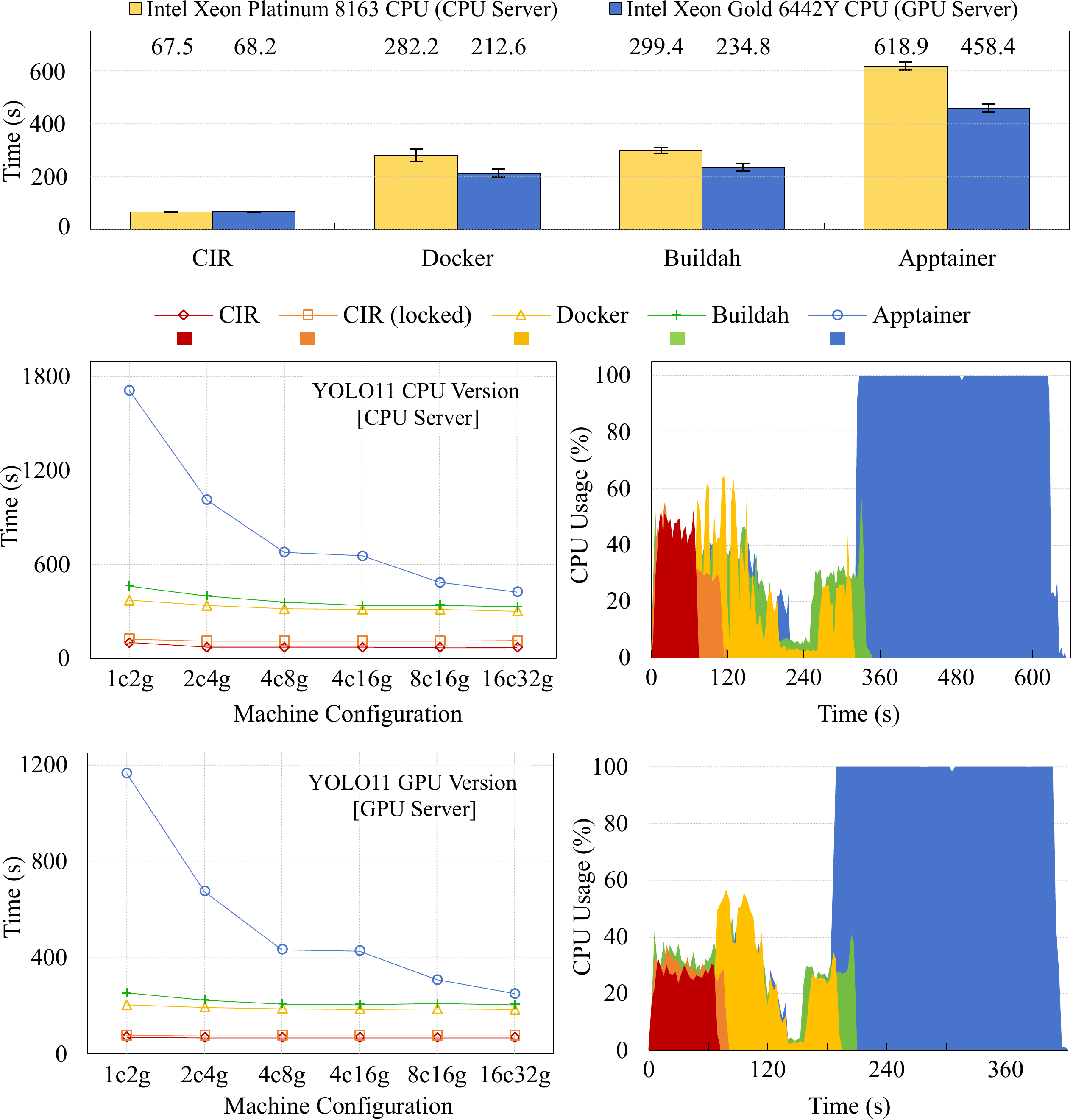}
    \caption{Build times of YOLO11 project under different number of CPU cores and memory sizes.}
    \Description{}
    \label{fig:cpu&mem}
\end{figure}

To assess the impact of computing resources on build time, we built YOLO11 on both CPU and GPU servers using a 500 Mbps bandwidth network. The upper part of Figure~\ref{fig:cpu&mem} shows the build times on the CPU server and the GPU server, both limited to 4 CPU cores and 16 GB of memory. We found that the CIR system was not only the fastest but also the most stable and the least affected by CPU performance. Although the GPU server’s CPU is more powerful than the CPU server’s, this difference translates to only about a 1\% improvement for the CIR system but over 20\% for the baseline systems. The variance in build time for our CIR system was approximately 0.2 s, compared to over 11 s for the baseline systems.

The lower-left panel of Figure~\ref{fig:cpu&mem} shows the build times on both servers characterized by varying numbers of CPU cores and memory sizes. Only the Apptainer system’s performance was significantly affected by the number of CPU cores. The lower-right portion of Figure~\ref{fig:cpu&mem} shows the reason by showing CPU usage throughout the build process. Apptainer’s use of SquashFS for image compression requires substantial CPU computation during the latter stages of the build, whereas the other systems kept CPU usage below 50\%.

\label{subsec:representative}
\begin{figure*}[ht]
    \centering
    \includegraphics[width=\linewidth]{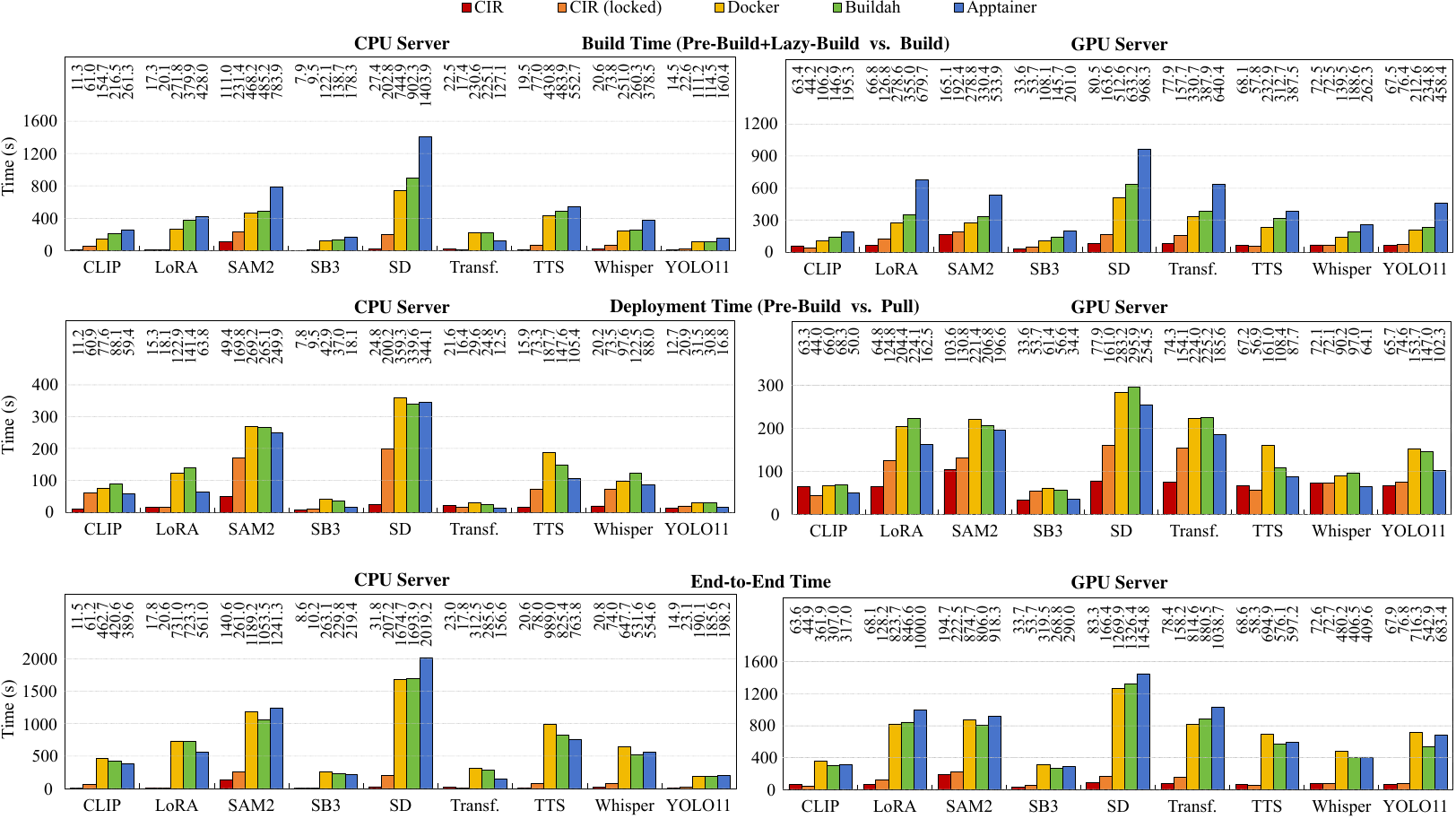}
    \caption{Build times for the benchmark suite on CPU and GPU servers, showing build time (\textit{top}), deployment time (\textit{middle}), and end-to-end time (\textit{bottom}). Both servers connect to the component server via a 500 Mbps network.}
    \Description{}
    \label{fig:representative}
\end{figure*}

We found that our CIR system consistently outperformed Docker and Buildah across scenarios. Figure~\ref{fig:representative} shows the detailed results for build time, deployment time, and end-to-end time of each project, using a representative hardware configuration (GPU server, 4 CPU cores, 16 GB of memory, and a 500 Mbps network bandwidth). We found that our CIR system achieves a reduction in build time of 77.26\% compared to Docker, 81.51\% compared to Buildah, and 86.94\% compared to Apptainer. In addition, our CIR system reduces the deployment time by 62.94\% relative to Docker, 61.25\% relative to Buildah, and 42.19\% relative to Apptainer. Critically, end-to-end time was reduced by 91.57\% compared to Docker, 90.58\% compared to Buildah, and 90.90\% compared to Apptainer.


\subsection{Storage Sharing}
\label{subsec:share}

Docker and Buildah store images at the \emph{layer} level, sharing identical layers across images to accelerate builds and pulls while reducing storage on both registries and deployment platforms. As shown in Table~\ref{tab:share}, storing all benchmark images individually requires 122.05 GB, whereas layer sharing reduces this to 113.90 GB (-6.7\%). It also shortens build time by 2.4\% (2616 s → 2553 s) and pull time by 6.8\% (1644 s → 1533 s).

\begin{table}[!htb]
\centering
\resizebox{\columnwidth}{!}{
\begin{tabular}{rrrrrrr}
\toprule[1.2pt]
\multirow{2}{*}{\textbf{}} & \multirow{2}{*}{\textbf{Layer}} & \multirow{2}{*}{\textbf{File}} & \multirow{2}{*}{\textbf{Chunk}} & \multirow{2}{*}{\textbf{\begin{tabular}[c]{@{}c@{}}Component\\ (passive)\end{tabular}}} & \multicolumn{2}{c}{\textbf{\begin{tabular}[c]{@{}c@{}}Component\\ (active)\end{tabular}}} \\ \cline{6-7} 
 &  &  &  &  & \multicolumn{1}{r}{\textbf{CPU}} & \textbf{GPU} \\

\midrule

Before (GB) & 122.05 & 107.78 & 102.62 & 100.02 & \multicolumn{1}{r}{17.98} & 54.24 \\
After (GB) & 113.90 & 91.11 & 82.20 & 78.40 & \multicolumn{1}{r}{9.66} & 16.53 \\
\textbf{Percentage} & \textbf{6.68\%} & \textbf{14.57\%} & \textbf{19.90\%} & \textbf{21.62\%} & \multicolumn{1}{r}{\textbf{46.27\%}} & \textbf{69.52\%} \\ 

\midrule

Before (Obj) & 255 & 614246 & 966273 & 3302 & \multicolumn{1}{r}{1957} & 2074 \\
After (Obj) & 244 & 312730 & 390609 & 992 & \multicolumn{1}{r}{784} & 799 \\
\textbf{Percentage} & \textbf{4.31\%} & \textbf{49.09\%} & \textbf{59.58\%} & \textbf{69.96\%} & \multicolumn{1}{r}{\textbf{59.94\%}} & \textbf{61.48\%} \\
\bottomrule[1.2pt]
\end{tabular}}
\caption{Applying different granularity sharing policy on benchmark suite. \textit{(Up)} Storage usage \textit{(Down)} Object count}
\label{tab:share}
\end{table}

\begin{figure}[htb]
    \centering
    \includegraphics[width=\linewidth]{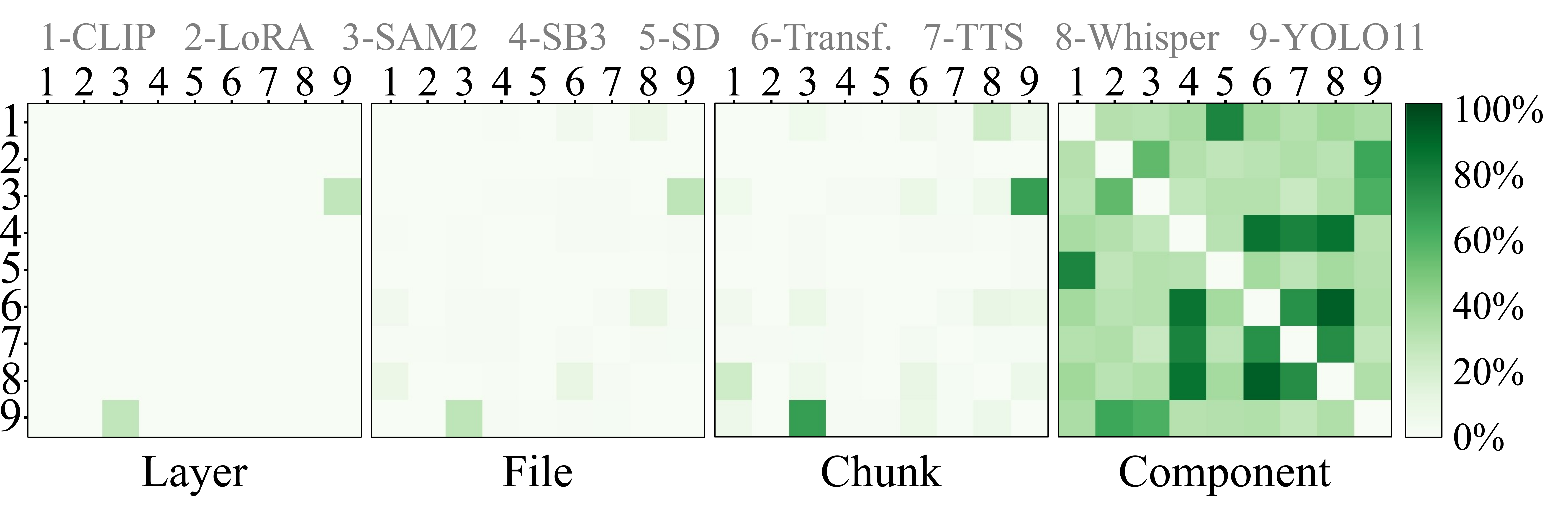}
    \vspace{-15pt}
    \caption{Pairwise sharing rate within benchmark suite.}
    \Description{}
    \label{fig:share}
\end{figure}



Layer-level sharing suffers from low reusability, as even small file changes invalidate entire layers~\cite{LayeredDeploy2021, LayeredDeploy2023, skourtis2019carving, HotC, Landlord2020, Landlord2023}. To improve the sharing rate, prior systems explored finer granularities: \emph{file-level} sharing (14.6\%) used by ORC~\cite{ORC}, Gear~\cite{Gear} and DupHunter~\cite{DupHunter2020, DupHunter2024}, and \emph{chunks-level} sharing (19.9\%) used by Slacker~\cite{Slacker} and Nydus~\cite{nydus}. However, finer granularity incurs heavy management overhead: in our benchmark suite, layer sharing required 255 objects, file-level 614k, and chunk-level 966k.

CIR adopts \emph{component-level} sharing, achieving 21.6\% with only 3302 objects. The sharing rate was slightly higher than that achieved at the chunk level, because these containers are slimmer than their Docker counterparts, having excluded download caches and Git project caches.


This sharing rate can be further increased through active sharing. Current sharing systems typically adopt a \emph{passive sharing method}, converting and analyzing images built by other systems without altering their content. In contrast, the CIR build system is able to employ an \emph{active sharing method} at component granularity, proactively selecting local reusable components that meet the specified dependencies. This approach raises the sharing rate to 46.27\% on a CPU server and 69.52\% on a GPU server. Additionally, the lazy-build time of all projects was reduced by 40.5\% (from 179.0 s to 106.6 s) on a CPU server and 68.2\% (from 622.4 s to 197.6 s) on a GPU server under the representative hardware configuration. Figure~\ref{fig:share} visualizes the pairwise sharing rates, which generally increase as granularity becomes finer, yet reach their highest value at the component level.
\section{Related Work}

Containers were initially adopted in cloud computing. Over time, their usage has expanded to high-performance computing (HPC)~\cite{gerhardt2017shifter, apptainer, Charliecloud} and edge devices~\cite{edgecontainer, TinyML}, thereby supporting a wide range of computing environments. As container images encapsulate both application code and execution environments, the efficiency of image construction and distribution has therefore become a critical factor in container-based deployment.

To accelerate image builds, prior work has proposed caching and reusing packages or images~\cite{FastBuild, update}, optimizing build strategies~\cite{buildpacks, img, kaniko, jib}, and reducing image size by multi-stage builds, removing redundant system software~\cite{distroless, Cimplifier}, or pruning unused files~\cite{SummSlim, slimtoolkit}. These approaches substantially reduce build time and image size during the development phase. However, they still construct platform-specific images on the development platform, limiting cross-platform reuse. Applying such build-time techniques at deployment time would incur higher latency than pulling pre-built images. In contrast, lazy-build enables deployment-time adaptation while achieving build latency even lower than image pulling.

Several recent systems explore deployment-time construction for containerized applications, such as XaaS~\cite{xaas} and coMtainer~\cite{comtainer}, but primarily target compiled-language workloads (e.g., C and C++) and focus on improving execution performance, whereas CIR targets interpreted-language workloads and optimizes both deployment performance and execution performance.

To improve container deployment performance, modern systems adopt lazy pulling technique. Lazy pulling prioritizes downloading only the essential parts of an image to start the container quickly, deferring the retrieval of non-essential parts to runtime. This approach can operate at the block level~\cite{DADI, Slacker, nydus}, file level~\cite{Starlight, FogDocker}, and image level~\cite{thalheim2018cntr}. CIR focuses on the generation of image and container file directories, while lazy pulling accelerates startup by introducing new container runtime file systems. These two lines of research are orthogonal and complementary. When combined, CIR and lazy pulling yield up to approximately 32\% additional performance improvement over existing methods.

\section{Discussion}

\paragraph{Compatibility with the Container Ecosystem}
CIR is integrated with mainstream container runtimes, including \texttt{containerd} and \texttt{CRI-O}, at the image management layer of the container stack. To support practical adoption, CIR enables automated migration of legacy container images by inferring environment declarations from existing software installations, allowing legacy artifacts to be converted into CIR specifications without semantic loss.

\paragraph{Implications for Existing Testing Pipelines}
Delegating parts of the build process to the deployment platform may affect established CI/CD testing pipelines. Because a single CIR may be deployed across heterogeneous platforms, the bar for validating final build outputs becomes higher. For applications that do not require cross-platform deployment, CIR does not introduce additional testing complexity compared to conventional workflows: the CIR can be lazily built on the target test platform, then after successful tests, all uniform components can be locked for that platform to guarantee subsequent deployments remain bit-for-bit consistent.

\paragraph{Cross-Platform Deployment: Costs and Benefits}
For applications that do require cross-platform deployment, testing complexity does increase: teams must verify that containers lazily built from the same CIR behave as expected on each target platform. CIR can record platforms that have passed validation and prevent builds on untested platforms by default. Nevertheless, as emphasized by emerging paradigms such as sky computing and edge computing, the flexibility and reach enabled by robust cross-platform execution can outweigh the additional testing overhead, delivering outsized benefits in portability, performance placement, and operational resilience.

\paragraph{Registry Feasibility} Unifying packages from multiple ecosystems has been demonstrated to be practical by \texttt{conda} and \texttt{Nix}, which convert heterogeneous package formats into their own formats and store them in their respective registries. These systems are designed for multi-version coexistence, consistency, and reproducibility, while CIR targets deployment performance. Currently, CIR intentionally focuses on \texttt{Debian}, \texttt{Python}, and \texttt{R}, which aligns with the ecosystem coverage prioritized by conda and covers the majority of machine learning and data analytics workloads. Due to the on-demand conversion strategy, the current uniform component repository uses about 1.6 TB of storage and serves the vast majority of needs. Fully converting all Debian packages adds 20 TB, and all Python packages add another 10 TB.

\paragraph{Extensibility} CIR is designed to be ecosystem-pluggable. Extending CIR to another interpreted language ecosystem requires four more components: versioning rule, package conversion, registry API adapter, and (optionally) application dependency analysis. In practice, each additional ecosystem introduces only about 3,000 LoC. Extending CIR to compiled languages would require an additional on-node compilation stage and warrants further investigation to avoid hurting cold-start performance.
\section{Conclusion}

In this paper, we proposed a novel CIR that functions as a lightweight and cross-platform container format. Unlike current image-building approaches that require developers to construct an execution environment for each platform, our CIR delegates this responsibility to the lazy-builder on the deployment platform. This approach enables the lazy-builder to choose the most suitable components based on the deployment platform's hardware and software configurations. Consequently, pre-built components can be fetched on demand, thereby enhancing reusability and reducing network usage. As a result, CIR facilitates cross-platform container deployment and seamless compute migration, making it well-suited for application migration in emerging computing paradigms such as sky computing, multi-cloud computing, and edge computing.
\bibliographystyle{ACM-Reference-Format}
\bibliography{ref}

@String{Computing = "Computing" }

@String{Computer = "{IEEE} Computer" }

@String{Academic = "Academic Press" }

@String{Springer = "Springer-Verlag" }

@inproceedings{comtainer,
author = {Gu, Yuhao and Chen, Haoquan and Chen, Xianjie and Du, Jiangsu and Chen, Zhiguang and Xiao, Nong and Zhang, Xianwei and Lu, Yutong},
title = {coMtainer: Compilation-assisted HPC Container Images with Enhanced Adaptability},
year = {2025},
isbn = {9798400714665},
publisher = {Association for Computing Machinery},
address = {New York, NY, USA},
url = {https://doi.org/10.1145/3712285.3759790},
doi = {10.1145/3712285.3759790},
booktitle = {Proceedings of the International Conference for High Performance Computing, Networking, Storage and Analysis},
pages = {586–601},
numpages = {16},
location = {
},
series = {SC '25}
}

@inproceedings{xaas,
author = {Copik, Marcin and Alnuaimi, Eiman and Kamatar, Alok and Hayot-Sasson, Valerie and Madonna, Alberto and Gamblin, Todd and Chard, Kyle and Foster, Ian and Hoefler, Torsten},
title = {XaaS Containers: Performance-Portable Representation With Source and IR Containers},
year = {2025},
isbn = {9798400714665},
publisher = {Association for Computing Machinery},
address = {New York, NY, USA},
url = {https://doi.org/10.1145/3712285.3759868},
doi = {10.1145/3712285.3759868},
booktitle = {Proceedings of the International Conference for High Performance Computing, Networking, Storage and Analysis},
pages = {533–555},
numpages = {23},
location = {
},
series = {SC '25}
}

@INPROCEEDINGS{update,
  author={Lu, Zhigang and Wu, Yuewen and Xu, Jiwei and Wang, Tao},
  booktitle={2019 IEEE International Conference on Fog Computing (ICFC)}, 
  title={An Acceleration Method for Docker Image Update}, 
  year={2019},
  pages={15-23},
  doi={10.1109/ICFC.2019.00010}}

@inproceedings{TinyML,
title={A VM Containerized Approach for Scaling TinyML Applications},
author={Meelis Lootus and Kartik Thakore and Sam Leroux and Geert Trooskens and Akshay Sharma and Holly Ly},
booktitle={Research Symposium on Tiny Machine Learning},
year={2021},
url={https://openreview.net/forum?id=m7Aqzt7Xf4S}
}

@INPROCEEDINGS{edgecontainer,
  author={Yang, Dali and Dai, Wenbin},
  booktitle={2022 IEEE 17th Conference on Industrial Electronics and Applications (ICIEA)}, 
  title={A Lightweight Container Design for Microservice-based Industrial Edge Applications}, 
  year={2022},
  pages={858-863},
  doi={10.1109/ICIEA54703.2022.10006175}}

@inproceedings{Charliecloud,
author = {Priedhorsky, Reid and Randles, Tim},
title = {Charliecloud: unprivileged containers for user-defined software stacks in HPC},
year = {2017},
isbn = {9781450351140},
publisher = {Association for Computing Machinery},
address = {New York, NY, USA},
url = {https://doi.org/10.1145/3126908.3126925},
doi = {10.1145/3126908.3126925},
booktitle = {Proceedings of the International Conference for High Performance Computing, Networking, Storage and Analysis},
articleno = {36},
numpages = {10},
location = {Denver, Colorado},
series = {SC '17}
}

@INPROCEEDINGS{SummSlim,
  author={Zhang, Zhicong and Huang, Heqing and Xu, Shaowen and Zhou, Qihang and Zhang, Tianshu and Jia, Xiaoqi and Zhang, Weijuan},
  booktitle={2024 IEEE 30th International Conference on Parallel and Distributed Systems (ICPADS)}, 
  title={SummSlim: A Universal and Automated Approach for Debloating Container Images}, 
  year={2024},
  pages={132-141},
  doi={10.1109/ICPADS63350.2024.00027}}

@inproceedings{Cimplifier,
author = {Rastogi, Vaibhav and Davidson, Drew and De Carli, Lorenzo and Jha, Somesh and McDaniel, Patrick},
title = {Cimplifier: automatically debloating containers},
year = {2017},
isbn = {9781450351058},
publisher = {Association for Computing Machinery},
address = {New York, NY, USA},
url = {https://doi.org/10.1145/3106237.3106271},
doi = {10.1145/3106237.3106271},
booktitle = {Proceedings of the 2017 11th Joint Meeting on Foundations of Software Engineering},
pages = {476–486},
numpages = {11},
location = {Paderborn, Germany},
series = {ESEC/FSE 2017}
}

@article{multicloud,
  author    = {Carlos Guerrero and Isaac Lera and Carlos Juiz},
  title     = {Resource optimization of container orchestration: a case study in multi-cloud microservices-based applications},
  journal   = {The Journal of Supercomputing},
  year      = {2018},
  volume    = {74},
  number    = {7},
  pages     = {2956--2983},
  doi       = {10.1007/s11227-018-2345-2},
  url       = {https://doi.org/10.1007/s11227-018-2345-2},
  issn      = {1573-0484},
  month     = jul,
  day       = {1}
}

@ARTICLE{edge1,
  author={Oleghe, Omogbai},
  journal={IEEE Access}, 
  title={Container Placement and Migration in Edge Computing: Concept and Scheduling Models}, 
  year={2021},
  volume={9},
  pages={68028-68043},
  doi={10.1109/ACCESS.2021.3077550}}

@article{edge2,
title = {Optimized container scheduling for data-intensive serverless edge computing},
journal = {Future Generation Computer Systems},
volume = {114},
pages = {259-271},
year = {2021},
issn = {0167-739X},
doi = {https://doi.org/10.1016/j.future.2020.07.017},
url = {https://www.sciencedirect.com/science/article/pii/S0167739X2030399X},
author = {Thomas Rausch and Alexander Rashed and Schahram Dustdar}
}

@misc{nydus,
  title        = {Nydus: Acceleration Framework for Container Image},
  author       = {Nydus Development Team},
  howpublished = {\url{https://nydus.dev/}},
  note         = {Accessed: 2024-12-10},
}

@misc{slimtoolkit,
  title        = {SlimToolkit: Optimize Your Experience with Containers},
  author       = {SlimToolkit Development Team},
  howpublished = {\url{https://github.com/slimtoolkit/slim}},
  year = {[n.\,d.]},
  note         = {Accessed: 2024-12-10},
}

@misc{tts,
  author       = {Eren, Gölge and
                  The Coqui TTS Team},
  title        = {Coqui TTS},
  month        = dec,
  year         = 2023,
  publisher    = {Zenodo},
  version      = {v0.22.0},
  doi          = {10.5281/zenodo.10363832},
  url          = {https://doi.org/10.5281/zenodo.10363832}
}

@inproceedings{transformers,
    title = "Transformers: State-of-the-Art Natural Language Processing",
    author = "Wolf, Thomas  and
      Debut, Lysandre  and
      Sanh, Victor  and
      Chaumond, Julien  and
      Delangue, Clement  and
      Moi, Anthony  and
      Cistac, Pierric  and
      Rault, Tim  and
      Louf, Remi  and
      Funtowicz, Morgan  and
      Davison, Joe  and
      Shleifer, Sam  and
      von Platen, Patrick  and
      Ma, Clara  and
      Jernite, Yacine  and
      Plu, Julien  and
      Xu, Canwen  and
      Le Scao, Teven  and
      Gugger, Sylvain  and
      Drame, Mariama  and
      Lhoest, Quentin  and
      Rush, Alexander",
    editor = "Liu, Qun  and
      Schlangen, David",
    booktitle = "Proceedings of the 2020 Conference on Empirical Methods in Natural Language Processing: System Demonstrations",
    month = oct,
    year = "2020",
    address = "Online",
    publisher = "Association for Computational Linguistics",
    url = "https://aclanthology.org/2020.emnlp-demos.6",
    doi = "10.18653/v1/2020.emnlp-demos.6",
    pages = "38--45",
}

@inproceedings{SAM2,
title={{SAM} 2: Segment Anything in Images and Videos},
author={Nikhila Ravi and Valentin Gabeur and Yuan-Ting Hu and Ronghang Hu and Chaitanya Ryali and Tengyu Ma and Haitham Khedr and Roman R{\"a}dle and Chloe Rolland and Laura Gustafson and Eric Mintun and Junting Pan and Kalyan Vasudev Alwala and Nicolas Carion and Chao-Yuan Wu and Ross Girshick and Piotr Dollar and Christoph Feichtenhofer},
booktitle={The Thirteenth International Conference on Learning Representations},
year={2025},
url={https://openreview.net/forum?id=Ha6RTeWMd0}
}

@article{sb3,
  author  = {Antonin Raffin and Ashley Hill and Adam Gleave and Anssi Kanervisto and Maximilian Ernestus and Noah Dormann},
  title   = {Stable-Baselines3: Reliable Reinforcement Learning Implementations},
  journal = {Journal of Machine Learning Research},
  year    = {2021},
  volume  = {22},
  number  = {268},
  pages   = {1--8},
  url     = {http://jmlr.org/papers/v22/20-1364.html}
}

@inproceedings{whisper,
author = {Radford, Alec and Kim, Jong Wook and Xu, Tao and Brockman, Greg and McLeavey, Christine and Sutskever, Ilya},
title = {Robust speech recognition via large-scale weak supervision},
year = {2023},
publisher = {JMLR.org},
booktitle = {Proceedings of the 40th International Conference on Machine Learning},
articleno = {1182},
numpages = {27},
location = {Honolulu, Hawaii, USA},
series = {ICML'23}
}

@INPROCEEDINGS {stablediffusion,
author = { Rombach, Robin and Blattmann, Andreas and Lorenz, Dominik and Esser, Patrick and Ommer, Bjorn },
booktitle = { 2022 IEEE/CVF Conference on Computer Vision and Pattern Recognition (CVPR) },
title = {{ High-Resolution Image Synthesis with Latent Diffusion Models }},
year = {2022},
pages = {10674-10685},
doi = {10.1109/CVPR52688.2022.01042},
url = {https://doi.ieeecomputersociety.org/10.1109/CVPR52688.2022.01042},
publisher = {IEEE Computer Society},
address = {Los Alamitos, CA, USA},
month =Jun}

@InProceedings{clip,
  title = 	 {Learning Transferable Visual Models From Natural Language Supervision},
  author =       {Radford, Alec and Kim, Jong Wook and Hallacy, Chris and Ramesh, Aditya and Goh, Gabriel and Agarwal, Sandhini and Sastry, Girish and Askell, Amanda and Mishkin, Pamela and Clark, Jack and Krueger, Gretchen and Sutskever, Ilya},
  booktitle = 	 {Proceedings of the 38th International Conference on Machine Learning},
  pages = 	 {8748--8763},
  year = 	 {2021},
  editor = 	 {Meila, Marina and Zhang, Tong},
  volume = 	 {139},
  series = 	 {Proceedings of Machine Learning Research},
  month = 	 {18--24 Jul},
  publisher =    {PMLR},
  url = 	 {https://proceedings.mlr.press/v139/radford21a.html},
}

@misc{overlayfs,
  author    = {Neil Brown},
  title     = {Overlay Filesystem},
  year      = {2019},
  howpublished = {\url{https://www.kernel.org/doc/Documentation/filesystems/overlayfs.rst}},
}

@inproceedings{lora,
title={Lo{RA}: Low-Rank Adaptation of Large Language Models},
author={Edward J Hu and Yelong Shen and Phillip Wallis and Zeyuan Allen-Zhu and Yuanzhi Li and Shean Wang and Lu Wang and Weizhu Chen},
booktitle={International Conference on Learning Representations},
year={2022},
url={https://openreview.net/forum?id=nZeVKeeFYf9}
}

@inproceedings {tensorflow,
author = {Mart{\'\i}n Abadi and Paul Barham and Jianmin Chen and Zhifeng Chen and Andy Davis and Jeffrey Dean and Matthieu Devin and Sanjay Ghemawat and Geoffrey Irving and Michael Isard and Manjunath Kudlur and Josh Levenberg and Rajat Monga and Sherry Moore and Derek G. Murray and Benoit Steiner and Paul Tucker and Vijay Vasudevan and Pete Warden and Martin Wicke and Yuan Yu and Xiaoqiang Zheng},
title = {{TensorFlow}: A System for {Large-Scale} Machine Learning},
booktitle = {12th USENIX Symposium on Operating Systems Design and Implementation (OSDI 16)},
year = {2016},
isbn = {978-1-931971-33-1},
address = {Savannah, GA},
pages = {265--283},
url = {https://www.usenix.org/conference/osdi16/technical-sessions/presentation/abadi},
publisher = {USENIX Association},
month = nov
}

@manual{amd,
  title        = {RDNA3 Shader Instruction Set Architecture},
  author       = {{Advanced Micro Devices}},
  organization = {Advanced Micro Devices, Inc.},
  year         = {2023},
  month        = {February},
  url          = {https://www.amd.com/content/dam/amd/en/documents/radeon-tech-docs/instruction-set-architectures/rdna3-shader-instruction-set-architecture-feb-2023_0.pdf},
}

@article{jetson,
title = {A Survey on optimized implementation of deep learning models on the NVIDIA Jetson platform},
journal = {Journal of Systems Architecture},
volume = {97},
pages = {428-442},
year = {2019},
issn = {1383-7621},
doi = {https://doi.org/10.1016/j.sysarc.2019.01.011},
url = {https://www.sciencedirect.com/science/article/pii/S1383762118306404},
author = {Sparsh Mittal},
}

@ARTICLE{edgecomputing,
  author={Shi, Weisong and Cao, Jie and Zhang, Quan and Li, Youhuizi and Xu, Lanyu},
  journal={IEEE Internet of Things Journal}, 
  title={Edge Computing: Vision and Challenges}, 
  year={2016},
  volume={3},
  number={5},
  pages={637-646},
  doi={10.1109/JIOT.2016.2579198}
}

@INPROCEEDINGS{npu,
  author={Esmaeilzadeh, Hadi and Sampson, Adrian and Ceze, Luis and Burger, Doug},
  booktitle={2012 45th Annual IEEE/ACM International Symposium on Microarchitecture}, 
  title={Neural Acceleration for General-Purpose Approximate Programs}, 
  year={2012},
  pages={449-460},
  doi={10.1109/MICRO.2012.48}
}

@article{tpu,
author = {Jouppi, Norman P. and Young, Cliff and Patil, Nishant and Patterson, David and Agrawal, Gaurav and Bajwa, Raminder and Bates, Sarah and Bhatia, Suresh and Boden, Nan and Borchers, Al and Boyle, Rick and Cantin, Pierre-luc and Chao, Clifford and Clark, Chris and Coriell, Jeremy and Daley, Mike and Dau, Matt and Dean, Jeffrey and Gelb, Ben and Ghaemmaghami, Tara Vazir and Gottipati, Rajendra and Gulland, William and Hagmann, Robert and Ho, C. Richard and Hogberg, Doug and Hu, John and Hundt, Robert and Hurt, Dan and Ibarz, Julian and Jaffey, Aaron and Jaworski, Alek and Kaplan, Alexander and Khaitan, Harshit and Killebrew, Daniel and Koch, Andy and Kumar, Naveen and Lacy, Steve and Laudon, James and Law, James and Le, Diemthu and Leary, Chris and Liu, Zhuyuan and Lucke, Kyle and Lundin, Alan and MacKean, Gordon and Maggiore, Adriana and Mahony, Maire and Miller, Kieran and Nagarajan, Rahul and Narayanaswami, Ravi and Ni, Ray and Nix, Kathy and Norrie, Thomas and Omernick, Mark and Penukonda, Narayana and Phelps, Andy and Ross, Jonathan and Ross, Matt and Salek, Amir and Samadiani, Emad and Severn, Chris and Sizikov, Gregory and Snelham, Matthew and Souter, Jed and Steinberg, Dan and Swing, Andy and Tan, Mercedes and Thorson, Gregory and Tian, Bo and Toma, Horia and Tuttle, Erick and Vasudevan, Vijay and Walter, Richard and Wang, Walter and Wilcox, Eric and Yoon, Doe Hyun},
title = {In-Datacenter Performance Analysis of a Tensor Processing Unit},
year = {2017},
issue_date = {May 2017},
publisher = {Association for Computing Machinery},
address = {New York, NY, USA},
volume = {45},
number = {2},
issn = {0163-5964},
url = {https://doi.org/10.1145/3140659.3080246},
doi = {10.1145/3140659.3080246},
journal = {SIGARCH Comput. Archit. News},
month = jun,
pages = {1–12},
numpages = {12}
}

@inproceedings{skycomputing,
author = {Stoica, Ion and Shenker, Scott},
title = {From cloud computing to sky computing},
year = {2021},
isbn = {9781450384384},
publisher = {Association for Computing Machinery},
address = {New York, NY, USA},
url = {https://doi.org/10.1145/3458336.3465301},
doi = {10.1145/3458336.3465301},
booktitle = {Proceedings of the Workshop on Hot Topics in Operating Systems},
pages = {26–32},
numpages = {7},
location = {Ann Arbor, Michigan},
series = {HotOS '21}
}

@inproceedings{imageSize,
author = {Durieux, Thomas},
title = {Empirical Study of the Docker Smells Impact on the Image Size},
year = {2024},
isbn = {9798400702174},
publisher = {Association for Computing Machinery},
address = {New York, NY, USA},
url = {https://doi.org/10.1145/3597503.3639143},
doi = {10.1145/3597503.3639143},
booktitle = {Proceedings of the IEEE/ACM 46th International Conference on Software Engineering},
articleno = {208},
numpages = {12},
location = {Lisbon, Portugal},
series = {ICSE '24}
}

@misc{cncf2023survey,
  title = {Cloud Native Computing Foundation Annual Survey 2023},
  author = {{Cloud Native Computing Foundation} and {The Linux Foundation}},
  year = {2023},
  url = {https://www.cncf.io/reports/cncf-annual-survey-2023/},
  note = {Accessed: 2024-11-08}
}

@article{dockerContainer,
author = {Merkel, Dirk},
title = {Docker: lightweight Linux containers for consistent development and deployment},
year = {2014},
issue_date = {March 2014},
publisher = {Belltown Media},
address = {Houston, TX},
volume = {2014},
number = {239},
issn = {1075-3583},
journal = {Linux J.},
month = mar,
articleno = {2}
}

@misc{ultralytics,
  author = {Jocher, Glenn and Qiu, Jing and Chaurasia, Ayush},
  title = {{Ultralytics YOLO}},
  year = {2023},
  month = jan,
  version = {8.0.0},
  url = {https://github.com/ultralytics/ultralytics},
  note = {AGPL-3.0 License}
}

@inproceedings {DupHunter2020,
author = {Nannan Zhao and Hadeel Albahar and Subil Abraham and Keren Chen and Vasily Tarasov and Dimitrios Skourtis and Lukas Rupprecht and Ali Anwar and Ali R. Butt},
title = {{DupHunter}: Flexible {High-Performance} Deduplication for Docker Registries},
booktitle = {2020 USENIX Annual Technical Conference (USENIX ATC 20)},
year = {2020},
isbn = {978-1-939133-14-4},
pages = {769--783},
url = {https://www.usenix.org/conference/atc20/presentation/zhao},
publisher = {USENIX Association},
month = {July}
}

@article{DupHunter2024,
author = {Zhao, Nannan and Lin, Muhui and Albahar, Hadeel and Paul, Arnab K. and Huan, Zhijie and Abraham, Subil and Chen, Keren and Tarasov, Vasily and Skourtis, Dimitrios and Anwar, Ali and Butt, Ali},
title = {An End-to-end High-performance Deduplication Scheme for Docker Registries and Docker Container Storage Systems},
year = {2024},
issue_date = {August 2024},
publisher = {Association for Computing Machinery},
address = {New York, NY, USA},
volume = {20},
number = {3},
issn = {1553-3077},
url = {https://doi.org/10.1145/3643819},
doi = {10.1145/3643819},
journal = {ACM Trans. Storage},
month = {June},
articleno = {18},
numpages = {35},
}

@article{DADI,
author = {Li, Huiba and Zhang, Zhihao and Yuan, Yifan and Du, Rui and Ma, Kai and Liu, Lanzheng and Zhang, Yiming and Hsu, Windsor},
title = {Block-level Image Service for the Cloud},
year = {2024},
issue_date = {February 2024},
publisher = {Association for Computing Machinery},
address = {New York, NY, USA},
volume = {20},
number = {1},
issn = {1553-3077},
url = {https://doi.org/10.1145/3620672},
doi = {10.1145/3620672},
journal = {ACM Trans. Storage},
month = {January},
articleno = {1},
numpages = {28},
}

@inproceedings {Slacker,
author = {Tyler Harter and Brandon Salmon and Rose Liu and Andrea C. Arpaci-Dusseau and Remzi H. Arpaci-Dusseau},
title = {Slacker: Fast Distribution with Lazy Docker Containers},
booktitle = {14th USENIX Conference on File and Storage Technologies (FAST 16)},
year = {2016},
isbn = {978-1-931971-28-7},
address = {Santa Clara, CA},
pages = {181--195},
url = {https://www.usenix.org/conference/fast16/technical-sessions/presentation/harter},
publisher = {USENIX Association},
month = {February},
}

@INPROCEEDINGS{HotC,
  author={Suo, Kun and Son, Junggab and Cheng, Dazhao and Chen, Wei and Baidya, Sabur},
  booktitle={2021 IEEE International Conference on Cluster Computing (CLUSTER)}, 
  title={Tackling Cold Start of Serverless Applications by Efficient and Adaptive Container Runtime Reusing}, 
  year={2021},
  pages={433-443},
  doi={10.1109/Cluster48925.2021.00018},
  ISSN={2168-9253},
  month={September},
}

@INPROCEEDINGS{Landlord2020,
  author={Shaffer, Tim and Hazekamp, Nicholas and Blomer, Jakob and Thain, Douglas},
  booktitle={2020 IEEE International Parallel and Distributed Processing Symposium (IPDPS)}, 
  title={Solving the Container Explosion Problem for Distributed High Throughput Computing}, 
  year={2020},
  pages={388-398},
  doi={10.1109/IPDPS47924.2020.00048},
  ISSN={1530-2075},
  month={May},
}

@ARTICLE{Landlord2023,
  author={Shaffer, Tim and Phung, Thanh Son and Chard, Kyle and Thain, Douglas},
  journal={IEEE Transactions on Parallel and Distributed Systems}, 
  title={Landlord: Coordinating Dynamic Software Environments to Reduce Container Sprawl}, 
  year={2023},
  volume={34},
  number={5},
  pages={1376-1389},
  doi={10.1109/TPDS.2023.3241598},
  ISSN={1558-2183},
  month={May},
}

@inproceedings {Starlight,
author = {Jun Lin Chen and Daniyal Liaqat and Moshe Gabel and Eyal de Lara},
title = {Starlight: Fast Container Provisioning on the Edge and over the {WAN}},
booktitle = {19th USENIX Symposium on Networked Systems Design and Implementation (NSDI 22)},
year = {2022},
isbn = {978-1-939133-27-4},
address = {Renton, WA},
pages = {35--50},
url = {https://www.usenix.org/conference/nsdi22/presentation/chen-jun-lin},
publisher = {USENIX Association},
month = {April}
}

@INPROCEEDINGS{FastBuild,
  author={Huang, Zhuo and Wu, Song and Jiang, Song and Jin, Hai},
  booktitle={2019 35th Symposium on Mass Storage Systems and Technologies (MSST)}, 
  title={FastBuild: Accelerating Docker Image Building for Efficient Development and Deployment of Container}, 
  year={2019},
  pages={28-37},
  doi={10.1109/MSST.2019.00-18},
  ISSN={2160-1968},
  month={May},
}

@inproceedings{FogDocker,
author = {Civolani, Lorenzo and Pierre, Guillaume and Bellavista, Paolo},
title = {FogDocker: Start Container Now, Fetch Image Later},
year = {2019},
isbn = {9781450368940},
publisher = {Association for Computing Machinery},
address = {New York, NY, USA},
url = {https://doi.org/10.1145/3344341.3368811},
doi = {10.1145/3344341.3368811},
booktitle = {Proceedings of the 12th IEEE/ACM International Conference on Utility and Cloud Computing},
pages = {51–59},
numpages = {9},
location = {Auckland, New Zealand},
series = {UCC'19}
}

@INPROCEEDINGS{Gear,
  author={Fan, Hao and Bian, Shengwei and Wu, Song and Jiang, Song and Ibrahim, Shadi and Jin, Hai},
  booktitle={2021 IEEE 41st International Conference on Distributed Computing Systems (ICDCS)}, 
  title={Gear: Enable Efficient Container Storage and Deployment with a New Image Format}, 
  year={2021},
  pages={115-125},
  doi={10.1109/ICDCS51616.2021.00020},
  ISSN={2575-8411},
  month={July},
}

@inproceedings {ORC,
author = {Vasily A. Sartakov and Llu{\'\i}s Vilanova and Munir Geden and David Eyers and Takahiro Shinagawa and Peter Pietzuch},
title = {{ORC}: Increasing Cloud Memory Density via Object Reuse with Capabilities},
booktitle = {17th USENIX Symposium on Operating Systems Design and Implementation (OSDI 23)},
year = {2023},
isbn = {978-1-939133-34-2},
address = {Boston, MA},
pages = {573--587},
url = {https://www.usenix.org/conference/osdi23/presentation/sartakov},
publisher = {USENIX Association},
month = {July},
}

@InProceedings{HeterogeneousHardware,
author="Gkikopoulos, Panagiotis
and Schiavoni, Valerio
and Spillner, Josef",
editor="Matos, Miguel
and Greve, Fab{\'i}ola",
title="Analysis and Improvement of Heterogeneous Hardware Support in Docker Images",
booktitle="Distributed Applications and Interoperable Systems",
year="2021",
publisher="Springer International Publishing",
address="Cham",
pages="125--142",
isbn="978-3-030-78198-9",
}

@InProceedings{Vulnerability,
author="Wist, Katrine
and Helsem, Malene
and Gligoroski, Danilo",
editor="Daimi, Kevin
and Arabnia, Hamid R.
and Deligiannidis, Leonidas
and Hwang, Min-Shiang
and Tinetti, Fernando G.",
title="Vulnerability Analysis of 2500 Docker Hub Images",
booktitle="Advances in Security, Networks, and Internet of Things",
year="2021",
publisher="Springer International Publishing",
address="Cham",
pages="307--327",
isbn="978-3-030-71017-0"
}

@article{TechnicalLag,
author = {Zerouali, Ahmed and Mens, Tom and Decan, Alexandre and Gonzalez-Barahona, Jesus and Robles, Gregorio},
title = {A multi-dimensional analysis of technical lag in Debian-based Docker images},
year = {2021},
issue_date = {Mar 2021},
publisher = {Kluwer Academic Publishers},
address = {USA},
volume = {26},
number = {2},
issn = {1382-3256},
url = {https://doi.org/10.1007/s10664-020-09908-6},
doi = {10.1007/s10664-020-09908-6},
journal = {Empirical Softw. Engg.},
month = {March},
numpages = {45}
}

@INPROCEEDINGS{LayeredDeploy2023,
  author={Zeng, Deze and Geng, Hongmin and Gu, Lin and Li, Zhexiong},
  booktitle={IEEE INFOCOM 2023 - IEEE Conference on Computer Communications}, 
  title={Layered Structure Aware Dependent Microservice Placement Toward Cost Efficient Edge Clouds}, 
  year={2023},
  pages={1-9},
  doi={10.1109/INFOCOM53939.2023.10229030},
  ISSN={2641-9874},
  month={May},
}

@INPROCEEDINGS{LayeredDeploy2021,
  author={Gu, Lin and Zeng, Deze and Hu, Jie and Jin, Hai and Guo, Song and Zomaya, Albert Y.},
  booktitle={IEEE INFOCOM 2021 - IEEE Conference on Computer Communications}, 
  title={Exploring Layered Container Structure for Cost Efficient Microservice Deployment}, 
  year={2021},
  pages={1-9},
  doi={10.1109/INFOCOM42981.2021.9488918},
  ISSN={2641-9874},
  month={May},
}

@inproceedings{skourtis2019carving,
  title={Carving perfect layers out of docker images},
  author={Skourtis, Dimitris and Rupprecht, Lukas and Tarasov, Vasily and Megiddo, Nimrod},
  booktitle={11th USENIX Workshop on Hot Topics in Cloud Computing (HotCloud 19)},
  year={2019}
}

@article{gerhardt2017shifter,
doi = {10.1088/1742-6596/898/8/082021},
url = {https://dx.doi.org/10.1088/1742-6596/898/8/082021},
year = {2017},
month = {oct},
publisher = {IOP Publishing},
volume = {898},
number = {8},
pages = {082021},
author = {Lisa Gerhardt and Wahid Bhimji and Shane Canon and Markus Fasel and Doug Jacobsen and Mustafa Mustafa and Jeff Porter and Vakho Tsulaia},
title = {Shifter: Containers for HPC},
journal = {Journal of Physics: Conference Series},
}

@misc{buildkit,
  author = {Moby},
  title = {BuildKit},
  howpublished = {\url{https://github.com/moby/buildkit}},
  year = {[n.\,d.]},
  note         = {Accessed: 2024-12-10},
}

@misc{apptainer,
  author = {LF Projects, LLC},
  title = {Apptainer: Application containers for Linux},
  howpublished = {\url{https://apptainer.org}},
  year = {[n.\,d.]},
  note         = {Accessed: 2024-12-10},
}

@misc{Buildah,
  author = {Containers organization},
  title = {Buildah: A tool that facilitates building OCI container images.},
  howpublished = {\url{https://buildah.io/}},
  year = {[n.\,d.]},
  note         = {Accessed: 2024-12-10},
}

@misc{buildpacks,
  author = {Cloud Native Computing Foundation},
  title = {Cloud Native Buildpacks},
  howpublished = {\url{https://buildpacks.io}},
  note         = {Accessed: 2024-12-10},
}

@misc{img,
  author = {genuinetools},
  title = {img: Standalone, daemon-less, unprivileged Dockerfile and OCI compatible container image builder.},
  howpublished = {\url{https://github.com/genuinetools/img}},
  year = {[n.\,d.]},
  note         = {Accessed: 2024-12-10},
}

@misc{distroless,
  author = {GoogleContainerTools},
  title = {"Distroless" Container Images},
  howpublished = {\url{https://github.com/GoogleContainerTools/distroless}},
  year = {[n.\,d.]},
  note         = {Accessed: 2024-12-10},
}

@inproceedings{thalheim2018cntr,
  title={CNTR: lightweight OS containers},
  author={Thalheim, J{\"o}rg and Bhatotia, Pramod and Fonseca, Pedro and Kasikci, Baris},
  booktitle={2018 USENIX Annual Technical Conference (USENIX ATC 18)},
  pages={199--212},
  year={2018}
}

@misc{conda,
  author = {The conda Organization},
  title = {conda: a cross-platform, language-agnostic binary package manager.},
  howpublished = {\url{https://conda.org}},
  year = {[n.\,d.]},
  note         = {Accessed: 2024-12-10},
}

@misc{nix,
  author = {Nix},
  title = {Nix: Declarative builds and deployments.},
  howpublished = {\url{https://nixos.org}},
  year = {[n.\,d.]},
  note         = {Accessed: 2024-12-10},
}

@misc{jib,
  author = {GoogleContainerTools},
  title = {Jib: Containerize your Java application.},
  howpublished = {\url{https://github.com/GoogleContainerTools/jib}},
  year = {[n.\,d.]},
  note         = {Accessed: 2024-12-10},
}

@misc{kaniko,
  author = {GoogleContainerTools},
  title = {kaniko - Build Images In Kubernetes.},
  howpublished = {\url{https://github.com/GoogleContainerTools/kaniko}},
  year = {[n.\,d.]},
  note         = {Accessed: 2024-12-10},
}

@inproceedings {skypilot,
author = {Zongheng Yang and Zhanghao Wu and Michael Luo and Wei-Lin Chiang and Romil Bhardwaj and Woosuk Kwon and Siyuan Zhuang and Frank Sifei Luan and Gautam Mittal and Scott Shenker and Ion Stoica},
title = {{SkyPilot}: An Intercloud Broker for Sky Computing},
booktitle = {20th USENIX Symposium on Networked Systems Design and Implementation (NSDI 23)},
year = {2023},
isbn = {978-1-939133-33-5},
address = {Boston, MA},
pages = {437--455},
url = {https://www.usenix.org/conference/nsdi23/presentation/yang-zongheng},
publisher = {USENIX Association},
month = apr
}

@inproceedings {skyplane,
author = {Paras Jain and Sam Kumar and Sarah Wooders and Shishir G. Patil and Joseph E. Gonzalez and Ion Stoica},
title = {Skyplane: Optimizing Transfer Cost and Throughput Using {Cloud-Aware} Overlays},
booktitle = {20th USENIX Symposium on Networked Systems Design and Implementation (NSDI 23)},
year = {2023},
isbn = {978-1-939133-33-5},
address = {Boston, MA},
pages = {1375--1389},
url = {https://www.usenix.org/conference/nsdi23/presentation/jain},
publisher = {USENIX Association},
month = apr
}

@article{hcontainer,
author = {Xing, Tong and Barbalace, Antonio and Olivier, Pierre and Karaoui, Mohamed L. and Wang, Wei and Ravindran, Binoy},
title = {H-Container: Enabling Heterogeneous-ISA Container Migration in Edge Computing},
year = {2022},
issue_date = {November 2021},
publisher = {Association for Computing Machinery},
address = {New York, NY, USA},
volume = {39},
number = {1–4},
issn = {0734-2071},
url = {https://doi.org/10.1145/3524452},
doi = {10.1145/3524452},
journal = {ACM Trans. Comput. Syst.},
month = jul,
articleno = {5},
numpages = {36}
}

@article{edgemigration,
author = {Machen, Andrew and Wang, Shiqiang and Leung, Kin K. and Ko, Bong Jun and Salonidis, Theodoros},
title = {Live Service Migration in Mobile Edge Clouds},
year = {2018},
issue_date = {February 2018},
publisher = {IEEE Press},
volume = {25},
number = {1},
issn = {1536-1284},
url = {https://doi.org/10.1109/MWC.2017.1700011},
doi = {10.1109/MWC.2017.1700011},
journal = {Wireless Commun.},
month = feb,
pages = {140–147},
numpages = {8}
}

@INPROCEEDINGS{heterogeneouscloud,
  author={Crago, Steve and Dunn, Kyle and Eads, Patrick and Hochstein, Lorin and Kang, Dong-In and Kang, Mikyung and Modium, Devendra and Singh, Karandeep and Suh, Jinwoo and Walters, John Paul},
  booktitle={2011 IEEE International Conference on Cluster Computing}, 
  title={Heterogeneous Cloud Computing}, 
  year={2011},
  volume={},
  number={},
  pages={378-385},
  doi={10.1109/CLUSTER.2011.49}
}

@article{Pubgrub,
title = {Conflict-driven answer set solving: From theory to practice},
journal = {Artificial Intelligence},
volume = {187-188},
pages = {52-89},
year = {2012},
issn = {0004-3702},
doi = {https://doi.org/10.1016/j.artint.2012.04.001},
url = {https://www.sciencedirect.com/science/article/pii/S0004370212000409},
author = {Martin Gebser and Benjamin Kaufmann and Torsten Schaub},
}
\end{document}